\documentclass[sigconf]{acmart}

\AtBeginDocument{%
  \providecommand\BibTeX{{%
    \normalfont B\kern-0.5em{\scshape i\kern-0.25em b}\kern-0.8em\TeX}}}

\setcopyright{acmcopyright}
\copyrightyear{2022}
\acmYear{2022}
\acmDOI{XXXXXXX.XXXXXXX}

\acmPrice{15.00}
\acmISBN{XXX-X-XXXX-XXXX-X/22/09}

\usepackage{tabularx}
\usepackage{multirow}
\usepackage{listings}
\usepackage{pgfplots}
\usepackage{pgfplotstable}
\usepackage{tkz-kiviat}
\usepackage{longtable}
\usepackage{stfloats}
\usepackage[font=footnotesize]{subfig}

\def\BibTeX{{\rm B\kern-.05em{\sc i\kern-.025em b}\kern-.08em
    T\kern-.1667em\lower.7ex\hbox{E}\kern-.125emX}}

\lstdefinestyle{defaultcodestyle}{
    basicstyle=\ttfamily\small,
    breakatwhitespace=false,
    breaklines=true,
    captionpos=b,
    keepspaces=true,
    numbers=left,
    numbersep=5pt,
    showspaces=false,
    showstringspaces=false,
    showtabs=false,
    tabsize=2
}
\usepgfplotslibrary{colorbrewer,groupplots}
\pgfplotsset{compat=1.16}
\pgfplotsset{every axis/.append style={
    title style={font=\scriptsize},
    label style={font=\scriptsize},
    tick label style={font=\scriptsize},
    legend style={font=\scriptsize}
}}
\pgfplotsset{
    cycle list/.define={my marks}{
        every mark/.append style={solid},mark=*\\
        every mark/.append style={solid},mark=square*\\
        every mark/.append style={solid},mark=triangle*\\
        every mark/.append style={solid},mark=diamond*\\
    },
}
\pgfplotsset{
    colormap/Set1-5,
    cycle multiindex* list={
        Set1-5
            \nextlist
        my marks
            \nextlist
        [3 of]linestyles
            \nextlist
        very thick
            \nextlist
    },
}
\pgfplotsset{/pgf/bar width=0.15cm}

\definecolor{Set1-5-1}{RGB}{228,26,28}
\definecolor{Set1-5-A}{RGB}{228,26,28}
\definecolor{Set1-5-2}{RGB}{55,126,184}
\definecolor{Set1-5-B}{RGB}{55,126,184}
\definecolor{Set1-5-3}{RGB}{77,175,74}
\definecolor{Set1-5-C}{RGB}{77,175,74}
\definecolor{Set1-5-4}{RGB}{152,78,163}
\definecolor{Set1-5-D}{RGB}{152,78,163}
\definecolor{Set1-5-5}{RGB}{255,127,0}
\definecolor{Set1-5-E}{RGB}{255,127,0}

\tikzstyle{linestyle1}=[Set1-5-1,very thick,solid,mark=*,every mark/.append style={solid}]
\tikzstyle{linestyle2}=[Set1-5-2,very thick,dashed,mark=square*,every mark/.append style={solid}]
\tikzstyle{linestyle3}=[Set1-5-3,very thick,dotted,mark=triangle*,every mark/.append style={solid}]
\tikzstyle{linestyle4}=[Set1-5-4,very thick,solid,mark=diamond*,every mark/.append style={solid}]
\tikzstyle{linestyle5}=[Set1-5-5,very thick,dashed,mark=*,every mark/.append style={solid}]
\tikzstyle{linestyle6}=[Set1-5-1,very thick,dotted,mark=square*,every mark/.append style={solid}]

\tikzstyle{barstyle1}=[Set1-5-1,fill=Set1-5-1,mark=none]
\tikzstyle{barstyle2}=[Set1-5-2,fill=Set1-5-2,mark=none]
\tikzstyle{barstyle3}=[Set1-5-3,fill=Set1-5-3,mark=none]
\tikzstyle{barstyle4}=[Set1-5-4,fill=Set1-5-4,mark=none]
\tikzstyle{barstyle5}=[Set1-5-5,fill=Set1-5-5,mark=none]

\newcommand\LegendImage[1]{
    \draw[
        /pgfplots/mesh=false,
        bar width=3pt,
        bar shift=0pt,
        mark repeat=2,
        mark phase=2,
        #1
    ] plot coordinates {
        (0cm,0cm)
        (0.3cm,0cm)
        (0.6cm,0cm)
    };
}

\newcommand\LegendEntry[1]{\node[anchor=west,black,font=\scriptsize,inner xsep=2pt]{#1};}

\newcommand*\circled[1]{\tikz[baseline=(char.base)]{\node[shape=circle,draw,inner sep=1pt] (char) {#1};}}

\begin{document}

\title{Secure Decentralized IoT Service Platform using Consortium Blockchain}

\author{Ruipeng Zhang}
\email{ruipeng-zhang@mocs.utc.edu}
\orcid{0000-0002-8180-5438}
\affiliation{%
  \institution{The University of Tennessee at Chattanooga}
  \streetaddress{615 McCallie Ave}
  \city{Chattanooga}
  \state{Tennessee}
  \country{USA}
  \postcode{423-425-5311}
}

\author{Chen Xu}
\email{kjx384@mocs.utc.edu}
\affiliation{%
  \institution{The University of Tennessee at Chattanooga}
  \streetaddress{615 McCallie Ave}
  \city{Chattanooga}
  \state{Tennessee}
  \country{USA}
  \postcode{423-425-5311}
}

\author{Mengjun Xie}
\email{mengjun-xie@utc.edu}
\orcid{0000-0001-5089-9614}
\affiliation{%
  \institution{The University of Tennessee at Chattanooga}
  \streetaddress{615 McCallie Ave}
  \city{Chattanooga}
  \state{Tennessee}
  \country{USA}
  \postcode{423-425-5311}
}

\renewcommand{\shortauthors}{Zhang and Xu, et al.}

\begin{abstract}
Blockchain technology has gained increasing popularity in the research of Internet of Things (IoT) systems in the past decade.
As a distributed and immutable ledger secured by strong cryptography algorithms, the blockchain brings a new perspective to secure IoT systems.
Many studies have been devoted to integrating blockchain into IoT device management, access control, data integrity, security, and privacy.
In comparison, the blockchain-facilitated IoT communication is much less studied.
Nonetheless, we see the potential of blockchain in decentralizing and securing IoT communications.
This paper proposes an innovative IoT service platform powered by consortium blockchain technology.
The presented solution abstracts machine-to-machine (M2M) and human-to-machine (H2M) communications into services provided by IoT devices.
Then, it materializes data exchange of the IoT network through smart contracts and blockchain transactions.
Additionally, we introduce the auxiliary storage layer to the proposed platform to address various data storage requirements.
Our proof-of-concept implementation is tested against various workloads and connection sizes under different block configurations to evaluate the platform's transaction throughput, latency, and hardware utilization.
The experiment results demonstrate that our solution can maintain high performance under most testing scenarios and provide valuable insights on optimizing the blockchain configuration to achieve the best performance.
\end{abstract}



\maketitle

\section{Introduction}
Integration of the IoT and blockchain began to bloom since distributed ledger technologies and cryptocurrencies.
Past IoT systems depend on centralized servers for communication and data storage, which often become the single point of risk to the security and privacy of the systems.
Blockchain technologies, however, enable collaboration between untrusted parties in a decentralized manner.
They eliminate the need for a trusted intermediary by creating a self-organized transaction network guided by a consensus protocol.
Thanks to the distributed network architecture, data on the blockchain has not only high availability but also strong integrity assured by cryptography algorithms and the immutable data structure.
As a result, blockchain technologies are widely applied to fields of IoT such as supply chain \cite{barolli_blockchain_2020}, power grid \cite{musleh_blockchain_2019}, healthcare \cite{ray_blockchain_2021}, and smart home \cite{abunaser_advanced_2019}.

However, prominent blockchain solutions have their performance, security, and privacy concerns.
Firstly, public permissionless blockchains such as Bitcoin and Ethereum oblige no restrictions on the participants who can create blocks and read transactions \cite{henry_blockchain_2018}.
In the context of IoT, the openness of such blockchains endangers user privacy and exposes IoT systems to cyber attacks.
In addition, the anonymity of permissionless blockchains makes it challenging to audit operations and trace in the network.
Secondly, to reach global state consistency in a trustless environment, public permissionless blockchains use costly consensus protocols like Proof of Work (PoW), Proof of Stake (PoS), or protocols that require particular hardware (e.g., Proof of Elapsed Time) \cite{salimitari_survey_2019}.
Such protocols do not apply to IoT systems where devices are heterogeneous and power-constrained.
They also cannot meet the throughput and latency requirements of IoT applications which often demand hundreds of transactions to be committed to the ledger within milliseconds to seconds.

Consortium blockchains, in comparison, remedy those disadvantages of public permissionless blockchains in a semi-trusted environment.
Unlike permissionless blockchains, a consortium blockchain is operated by a group of collaborating entities, and only authorized nodes of these entities can commit blocks to the ledger \cite{dib2018consortium}.
On a consortium blockchain, the ledger can also be made visible to all members or part of the group via access control policies.
Since participating authorities manage blockchain actors, it is more feasible to trace transaction flows and construct audit trails.
Regarding consensus protocols, consortium blockchains assume that transaction validators are predefined and semi-trusted.
Less resource-demanding consensus protocols such as Practical Byzantine Fault Tolerance (PBFT) \cite{castro1999practical} and Raft \cite{ongaro2014search} can be utilized in consortium blockchains to improve scalability.
Therefore, consortium blockchains usually yield much better performance than public permissionless blockchains \cite{bamakan_survey_2020}.

We have reviewed existing IoT and blockchain integration in literature and discovered a research gap in realizing decentralized secure and scalable M2M and H2M communication with consortium blockchain technology.
That is, overcoming the downsides of public permissionless blockchains for IoT communications by replacing them with consortium blockchains.
A few research groups \cite{Ali2018}\cite{Hang2019}\cite{pajooh_hyperledger_2021} explored the application of permissioned consortium blockchains and smart contracts in securing IoT communications and sensor data.
However, they either are very focused on a specific blockchain application of IoT, lack a robust system design, or miss an extensive evaluation of their proposed approach.
There is a strong need for a more generalized, clearly-defined, and well-tested framework for IoT communications utilizing the consortium blockchain technology.

In this paper, we present a decentralized service platform for secure M2M and M2H communications inside an IoT environment based on a permissioned consortium blockchain.
Instead of implementing a specific application of IoT with blockchain, the proposed work aims to establish a generic communication framework for various IoT systems.
The consortium blockchain is a secure and scalable communication channel for IoT devices and applications in our solution.
The communication protocol is formulated as services defined by IoT devices and provided to applications.
Meanwhile, exchanging messages become blockchain transactions and are conveyed through the blockchain network. 
In addition, optional auxiliary storage is introduced to fit the proposed framework into diverse application scenarios, such as sensor data archives and real-time messaging.
Finally, the framework provides a lightweight software development kit (SDK) and platform gateways to simplify blockchain operations for resource-constrained IoT devices and application developers.
In contrast to related studies, the proposed framework has the following advantages:

1) \textit{Generality and versatility}: Our framework is designed to power a wide range of IoT applications.
It abstracts the communication protocol into services that IoT devices can customize.
It also imposes minimum assumptions about the underlying blockchain features or storage types.
For example, it does not rely on a specific function offered by a particular blockchain platform or use a dedicated storage solution exclusively.
Therefore, it can support various IoT applications to satisfy communication and data processing requirements.
    
2) \textit{Interoperability}: The proposed solution can be integrated into existing IoT systems smoothly.
With the help of straightforward SDKs and platform gateways, an application developer can not only bring new IoT devices to the proposed framework but also migrate legacy IoT devices or systems easily.
Meanwhile, our solution works well with existing IoT identity management and access control infrastructure and can reuse digital identities already in place.
    
3) \textit{High performance}: As illustrated by the performance evaluation results, our framework has considerable read and write throughput and reasonable latency even under high workloads.
Our solution works much more efficiently for semi-trusted consortium environments than permissionless blockchain-based IoT communication solutions, which rely on heavy consensus protocols.

Our contributions to this work are as follows:

1) We have proposed a novel consortium blockchain-based IoT platform for secure and decentralized IoT communications.
The platform models IoT communications as services powered by smart contracts and blockchain transactions.
In this paper, we elaborate on the system design and implementation that make the platform successful for supporting diverse IoT applications.
    
2) We have evaluated the proposed platform extensively to demonstrate its performance under various workloads and block configurations.
The experiment results illustrate that the platform can achieve high read and write throughput and latency for most workloads.
This work also discusses how our solution can address security and privacy concerns.

3) Use case studies are presented to showcase the versatility of our solution.
Additionally, we open source the platform, SDKs, use case demos, and testbed set-up scripts to promote the reproducibility of this research.

The remainder of this paper is organized as follows:
Section~\ref{sec:related_work} reviews related work of blockchain-based IoT identity management and access control, data storage and marketplace, and device manipulation.
Section~\ref{sec:system_design} describes the proposed IoT service platform's architecture and key processes.
Section~\ref{sec:implementation} briefly presents the proof of concept platform implementation and several use case studies.
Then, we evaluate the performance of the proposed platform with a series of experiments and explore its security and privacy implications in Section~\ref{sec:evaluation}.
Finally, Section~\ref{sec:conclusion} concludes this paper and discusses future research directions.
\section{Related Work}
\label{sec:related_work}

The integration of blockchain and IoT has been extensively studied since the emergence of blockchain technology.
Attempts have been made to address challenges of IoT, such as distributed and heterogeneous architecture, device, data security, and profitability utilizing blockchains \cite{Panarello2018}.
In this section, we will review the current development of IoT blockchain integration in three specific areas related to our proposed service platform: identity management and access control, data storage and marketplace, and device command and control.

\subsection{Identity Management and Access Control}

Identity management (IdM) and access control are keys to IoT device and data security and trust.
It is difficult to apply traditional IdM systems to IoT environments, especially distributed and collaborative ones, due to their centralized nature, security vulnerabilities, and service fragmentation \cite{Zhu2018}.
IoT's security, scalability, and interoperability requirements call for new IdM and access control paradigms, and blockchain-based solutions are promising answers.

One approach to building a blockchain-based IdM is recreating public key infrastructure (PKI) using blockchain and smart contracts \cite{fromknecht2014certcoin} \cite{axon2015privacy} \cite{al2017scpki} \cite{Bouras2021}.
The blockchain-based PKI supports the same critical operations like registration, verification, and revocation as a traditional centralized PKI, with improvements in security and privacy.
Other research takes a different approach by building identity systems tailored to specific blockchain implementation.
For instance, Sovrin \cite{reed2016technical} and its underlying Hyperledger Indy blockchain\cite{hyperledger_indy_2020} provide a full-stack solution to decentralized, self-sovereign IdM on a public permissioned blockchain.
Finally, storing identities off-chain (e.g., in traditional PKI) and linking them back to the blockchain is also discussed and utilized in projects such as Hyperledger Fabric \cite{Androulaki2018}.

Once identity management has been established for an IoT system, one can further introduce access control to IoT data and regulate the communications between devices.
Traditional access control methods, including role-based access control (RBAC), attribute-based access control (ABAC), and capability-based access control (CBAC) are less capable of supporting the enormous, heterogeneous, and decentralized IoT environments.
Considering how access decisions are made, existing blockchain-based access control methods can be categorized into 1) transaction-based access control and 2) smart contract-based access control \cite{Riabi2019}.
Transaction-based access control methods such as FairAccess \cite{ouaddah2016fairaccess} leverage blockchain as an immutable, distributed storage for access tokens, while the generation and verification of those tokens take place off-chain.
On the other hand, smart contract-based access control focuses more on decentralizing the decision-making process with smart contracts.
For example, IoTChain \cite{alphand2018iotchain} lets resource owners define smart contracts for granting client access and generating access tokens.

\subsection{Data Storage and Marketplace}

The fast advance in distributed ledger technology invites new opportunities for distributed data storage, data sharing, and data monetization.
Blockchain is a distributed, immutable database system where IoT data like sensor readings and access logs may be stored.
However, due to the block size limitation and scalability considerations, storing IoT data off-chain is more practical.
Thus, hybrid blockchain-based storage networks have been proposed to reduce the chances of single point of failures (SPOF) as seen in traditional centralized storage systems, provide data integrity and security, and lower the storage cost \cite{Nazanin2020}.

For instance, general-purpose blockchain-based solutions that support bulk data storage, such as Storj \cite{wilkinson2014storj}, Sia \cite{vorick2014sia}, and FileCoin \cite{filecoin}, store only metadata of data blocks on the blockchain to prevent the bloating issue.
On such platforms, files are split into smaller blocks and sent to the underlying distributed storage network composed of miners or storage nodes.
To encourage participation and prevent dishonest behavior and free-riding, they also introduce new consensus protocols and cryptocurrencies to compensate for miners' storage and bandwidth usage.

The rise of blockchain technology and IoT also accelerates the growth of the market of IoT data.
With blockchain technology, an IoT data marketplace can become fully decentralized and autonomous, while reducing cost, improving transaction efficiency, and promoting data privacy.
Research in this area has been centered around ensuring data authenticity and provenance, secure data transfer, and payment processing \cite{Ozyilmaz2018} \cite{Tzianos2019} \cite{Bajoudah2019} \cite{Lin2020}.
Additionally, industry-led initiatives such as IOTA \cite{iota}, XBR \cite{xbr}, and Streamr \cite{Streamr2017} also provide real-world insights on the monetization of IoT data on the blockchain.
\subsection{Device Manipulation}

Blockchain technology also enables fully decentralized M2M and H2M communications for IoT systems.
With smart contracts, a blockchain-based IoT system is capable of autonomous decision-making based on business logic \cite{Panarello2018}.
Such systems are relieved from centralized device management and control, hence less vulnerable to the single point of failure.
Meanwhile, the blockchain and smart contract ensure communication integrity and provide an immutable audit trail.

Slock.it \cite{slockit} is among the first applications to leverage blockchain and smart contracts for controlling embedded devices.
It envisages a blockchain-based economy of things where people can rent their unused assets, such as bikes, to others through smart contracts.
A user can discover and lease assets on the Slock.it platform, and the whole process is enabled by smart asset controllers or IoT devices connected to the Ethereum network.
Upon successful payment, an asset will be unlocked by its controller as instructed by smart contracts.

Apart from securely sharing physical assets, Ethereum has been used as a decentralized M2M channel for IoT, thanks to its popularity and versatility.
{\protect\NoHyper\citeauthor{Fakhri2018}\protect\endNoHyper} \cite{Fakhri2018} proposed a proof-of-concept demonstration on replacing MQTT \cite{mqtt_2022} with Ethereum blockchain for communications between IoT devices.
The devices can talk to each other by reading data from and writing data to the blockchain via a smart contract intermediary.
{\protect\NoHyper\citeauthor{Wickstrom2021}\protect\endNoHyper} \cite{Wickstrom2021} introduced an Ethereum-based protocol for IoT device management and task handling.
This protocol utilizes two smart contracts to register an IoT device and create tasks for it.
The authors pointed out that their protocol could reduce network attack risks of IoT devices because they can use the blockchain as a secure channel for remote command and control while disallowing other incoming network connections.

Finally, several studies have implemented decentralized IoT device management and communication using permissioned blockchain framework \cite{Ali2018}\cite{Hang2019}.
Compared to permissionless blockchains, a permissioned blockchain inherently integrates with identity management and authentication so that the participation of the consensus can be verified and authorized.
It usually comes with better scalability and energy efficiency because its consensus can be achieved without computationally expensive mining.
{\protect\NoHyper\citeauthor{Ali2018}\protect\endNoHyper} \cite{Ali2018} discussed a Hyperledger Fabric-based IoT architecture for the smart home scenario where every device stores and shares its data via blockchain transactions.
In such an environment, a smart device can request services from other devices by communicating directly or indirectly through the cloud.
A smart contract also guards the list of devices and their shared secret keys.
{\protect\NoHyper\citeauthor{Hang2019}\protect\endNoHyper} \cite{Hang2019} outlined a blockchain platform for securing IoT sensing data integrity.
The platform provides smart contracts for registering and querying IoT devices and creating and deploying tasks that IoT devices can process.
In the end, a device owner can receive notifications about the events generated by the tasks from the blockchain.
The authors also detailed the implementation of their proposed platform that utilizes Hyperledger Fabric.

\section{System Design}
\label{sec:system_design}

\begin{figure}[htb]
\centering
\includegraphics[width=\linewidth]{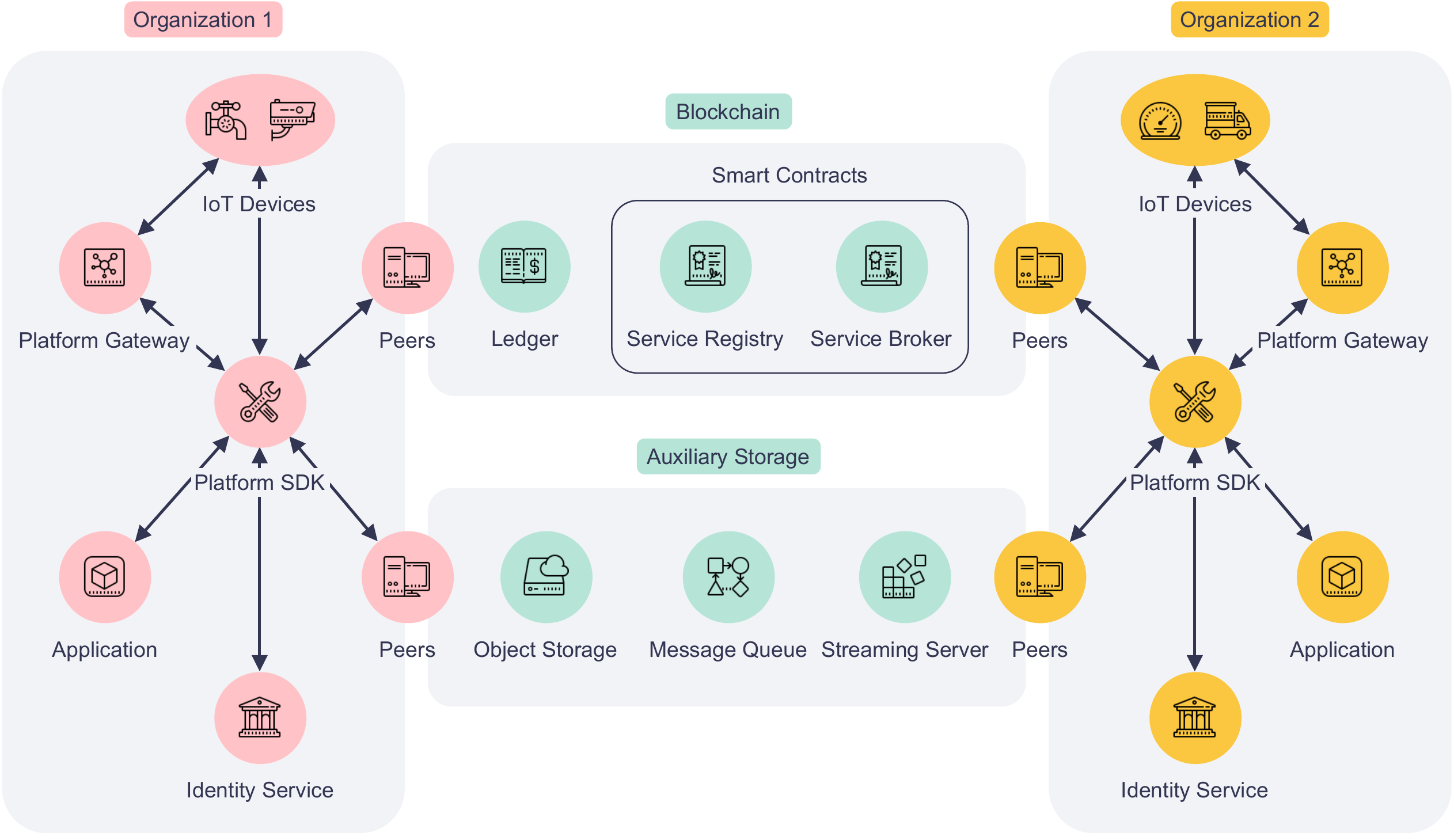}
\caption{IoT service platform architecture.\label{fig:arch}}
\end{figure}

An architectural overview of our proposed IoT service platform is illustrated in Figure~\ref{fig:arch}.
This overview shows the proposed platform for a consortium composed of two organizations for demonstration purpose.
At its core, the platform is powered by a consortium blockchain.
Peer nodes from two organizations together serve the distributed ledger and smart contracts.
These nodes also execute smart contracts when requested by the blockchain actors, which can be IoT devices or applications, and endorse blockchain transactions.
IoT devices from any organization within the consortium may connect to the platform directly or through the platform gateway that handles blockchain operations on the device's behalf.
To facilitate the integration of various IoT devices and accelerate application development, we also incorporate an SDK to reduce the complexity for devices and applications to interact with the platform.
At least one identity service is required to provide digital identities to each IoT device, application, and peer node in every organization of the consortium, although multiple organizations may share the same identity service.
Finally, auxiliary storage is introduced to enable data sharing between different organizations.
The actual storage types and implementations are affected by individual application requirements.

The following are core components of the proposed platform:

1) \textit{IoT Devices}: IoT devices connect the physical space to cyberspace.
On our platform, they are the primary service providers.
Services provided by IoT devices either measure or control their environment, given that the device is a sensor or actuator.
For example, a thermometer service reports the temperature of its surroundings, while a thermostat provides the service of adjusting the room temperature.
An IoT device may connect to the blockchain directly or through a platform gateway if its computational resources are constrained.

2) \textit{Platform Gateway}: A platform gateway bridges the communication between IoT devices and the platform.
Often, an IoT device is not able to participate in blockchain transactions directly due to energy or computational power constraints or because it is not programmable.
In this case, it can delegate blockchain operations to the platform gateway without changing its inherent communication protocols.

3) \textit{Consortium Blockchain}: At the core of the proposed platform, a consortium blockchain is employed as the distributed ledger that records all IoT devices and their services.
It is also the primary communication channel between IoT services and their consumers, offering much better performance and scalability compared to permission-less public blockchains.
The IoT network can become decentralized, meaning there are no longer centralized servers that are often SPOF.
Moreover, the platform can leverage access control features introduced by consortium blockchains to secure IoT services from unauthorized access.
Finally, since all changes to the IoT services and the service request and responses are recorded immutably on the ledger, the blockchain can essentially serve as a data historian for data auditing.

4) \textit{Peer}: Peers, or peer nodes, are computers that perform blockchain operations or offer auxiliary storage to IoT devices and applications of the platform.
When serving as blockchain peers, they are responsible for hosting the distributed ledger and executing smart contracts.
They are also essential in transaction endorsement.
As for the auxiliary storage, a peer can be tailored to specific application needs.
For instance, it can be a distributed storage network storage node, a distributed message queue broker, or a proxy server of media streams.
Any organization in the consortium can contribute peers to the network.
It is crucial to ensure a fair amount of peers from different organizations to achieve meaningful decentralization, and peers should be placed close to their neighboring IoT devices and gateways to reduce network overhead.

5) \textit{Service Registry \& Service Broker}: Service registry and broker are smart contracts that manage IoT devices and services as well as process service requests and responses.
During device and service registration, the service registry updates the ledger with the latest device or service information provided by the IoT device.
The service registry also provides interfaces for querying any IoT devices and services.
When invoked, the service broker smart contract will append service requests and responses to the ledger.
IoT devices and applications observe and respond to ledger updates, eventually achieving asynchronous communications between service providers and consumers.

6) \textit{Platform SDK}: Platform SDK provides an application programming interface (API) of the platform to IoT service and application developers.
It encapsulates functions for registering devices and services, sending service requests and responses, querying platform data, etc.
The goal of platform SDK is to conceal the complexity of blockchain operations from IoT devices and application developers.

7) \textit{Application}: An application interacts with IoT devices via the services published on the platform.
For example, an application can be an industrial control system (ICS) that monitors sensor readings or a smartphone app that displays room temperature and security camera feed in a smart home environment.
Additionally, applications can provide services to other IoT devices and applications on the platform.

8) \textit{Identity Service}: Every participating actor of the proposed platform, such as an IoT device, an application, or a peer node, is recognized by its digital identity.
Therefore, we need an identity service for every organization to issue, renew, and revoke those identities.
The platform allows organizations of the consortium to employ their own identity services so that each organization has complete control over its assets, such as peers, IoT devices, and applications.
In addition, the consortium blockchain can enforce access control to IoT services through a set of policies defined for digital identities.

9) \textit{Auxiliary Storage}: Although IoT devices and applications mainly communicate through the blockchain, it is sometimes desirable for IoT devices to share data off-chain.
The proposed platform encompasses an optional auxiliary storage system to satisfy various data storage needs.
For example, the auxiliary storage can be a distributed object storage system that stores historical humidity values, a distributed message queue for sharing real-time data like PM2.5 readings, or a proxy for streaming real-time binary data like a camera feed.
The uniform resource identifier (URI) can be passed to the data consumer via IoT services.
Finally, the data may be encrypted using the data consumer's identity to provide confidentiality.

The following sections will elaborate on the necessary procedures of the IoT service platform and explain how our solution takes advantage of the consortium blockchain and auxiliary storage to secure IoT systems.
\subsection{Device \& Service Registration}
\label{sec:device-service-registration}

\begin{figure}[htb]
\centering
\includegraphics[width=\linewidth]{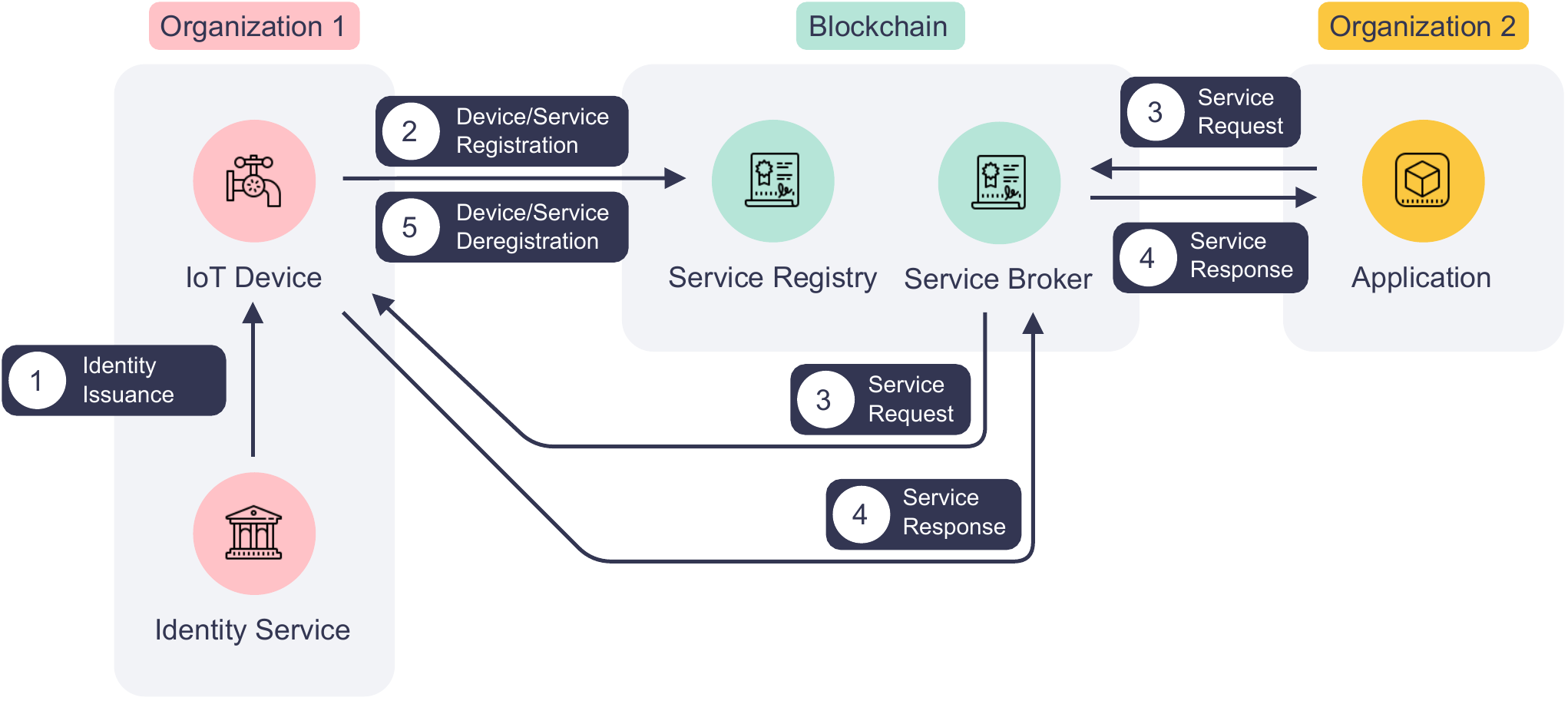}
\caption{The lifecycle of an IoT device and its service.\label{fig:service-lifecycle}}
\end{figure}

Figure~\ref{fig:service-lifecycle} depicts the lifecycle of an IoT device and its service on the proposed platform.
To expose functions to the network, an IoT device first must register itself and its services on the platform, as marked by step \circled{1} and \circled{2} in Figure~\ref{fig:service-lifecycle}.
Before the registration process, the device obtains its digital identity from the identity service of its belonging organization.
Then, it announces the device information and services by invoking the service registry smart contract.
If successfully validated, the device and service information will be stored on the blockchain.
The registration process is required because device and service information is crucial to the authentication and access control processes, as described in Section~\ref{sec:authentication-access-control}.

\begin{table*}[htbp]
\caption{Device and service registration information.\label{tbl:device-service-registration-information}}
\begin{tabularx}{\textwidth}{m{1.25in}m{1.25in}l}
\toprule
Category                     & Field   & Description                                                       \\
\midrule
\multirow{5}{*}{Device Registration}  & Device ID        & Device identifier. Must be unique within an organization.                  \\
                                      & Organization ID  & ID of the organization to which the device belongs.                        \\
                                      & Name             & Nickname of the device.                                                    \\
                                      & Description      & A short summary of the device's function and usage.                        \\
                                      & Last Update Time & The date and time when the last update is made to the device information.  \\
\midrule
\multirow{6}{*}{Service Registration} & Name             & Name of the service. Must be unique among all services of the same device. \\
                                      & Device ID        & Identifier of the device which provides the service.                       \\
                                      & Organization ID  & ID of the organization to which the device belongs.                        \\
                                      & Version          & Version number of the service.                                             \\
                                      & Description      & A short summary of the service's function and usage.                       \\
                                      & Last Update Time & The date and time when the last update is made to the service information. \\
\bottomrule
\end{tabularx}
\end{table*}

1) \textit{Device Registration}:
The first step for an IoT device to be registered with the proposed platform is receiving its digital identity from the identity service provider of its organization.
Unlike many other blockchain-based IoT platforms in the literature that employ custom device identity and registration processes, our platform can reuse existing digital identities issued by the organization's PKI.
Thus, organizations can follow the standard procedure of signing and issuing identity certificates using an off-the-shelf certificate authority (CA).
Not only can the platform benefit from the robust security of a PKI, but also participating organizations can integrate their IoT device management system easily with the proposed platform.
Consequently, IoT devices can be decommissioned by deregistering the service from the service registry and putting their identity certificate in the Certificate Revocation List (CRL).

Then, the IoT device can register itself by invoking the service registry with the required information.
Table~\ref{tbl:device-service-registration-information} describes the information required by the service registry smart contract to register a device and service to the platform.
The most important fields that must be provided by an IoT device are its device ID and the organization ID, which are used to locate a specific device on the platform.
The device ID is generated from the IoT device's digital certificate, while the organization ID is usually the name of an organization's identity service.
Other information, such as name, description, and last update time, are human-readable metadata for platform users.

2) \textit{Service Registration}:
Registered IoT devices publish their services to the network via the service registry smart contract.
Similar to device registration, an IoT device is required to provide service information to the service registry for each service, as described in the bottom half of Table~\ref{tbl:device-service-registration-information}.
Such a service can be measuring the room's humidity or setting the refrigerator's temperature.
An IoT device may declare any number of services on the blockchain by repeating the registration process.
The service name, device ID, and organization ID identify a unique service in the network.
Therefore, the service name should be unique among all services of a given device.
The service version number and last update time are useful fields for application developers and system auditors to keep track of the service updates.
The description field offers a summary of the service and its usage.
In accordance with the device decommission process, the service registry also allows IoT devices to deregister their services by calling the corresponding function in the smart contract, as shown in step \circled{5} in Figure~\ref{fig:service-lifecycle}.

\subsection{Authentication \& Access Control}
\label{sec:authentication-access-control}

\begin{figure}[t]
\centering
\includegraphics[width=\linewidth]{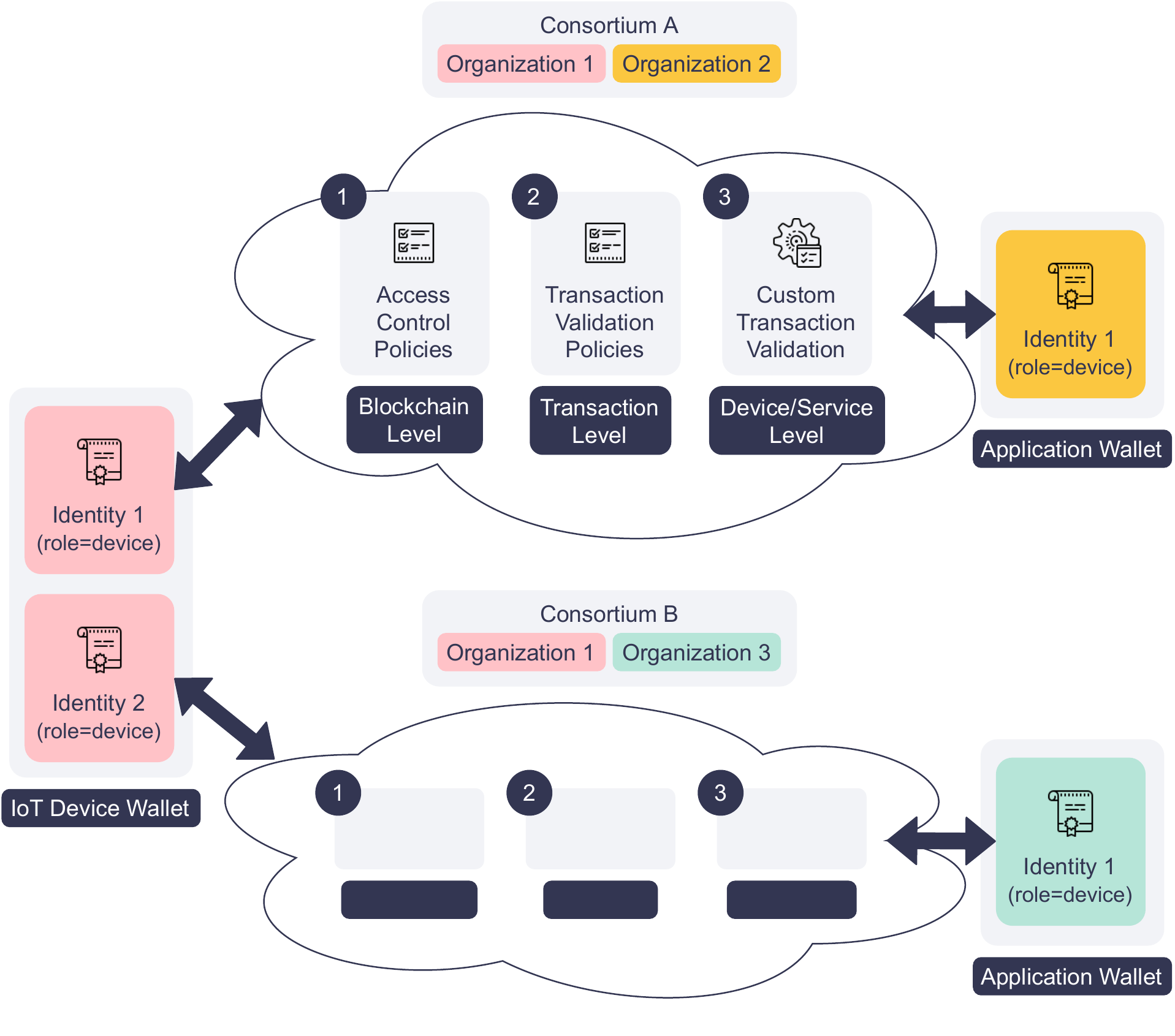}
\caption{The proposed multi-layered access control model.\label{fig:access-control}}
\end{figure}

Our proposed platform enforces authentication and access control to prevent unauthorized access.
As stated in Section~\ref{sec:device-service-registration}, the platform's authentication and access control rely on digital identities issued by the identity service.
All platform actors, such as IoT devices and applications, must have at least one valid digital identity to interact with the blockchain.
The smart contracts will verify the identity of calling actors to see if they have permission to read or write the distributed ledger.
To enable cross-organization operations, the actors can possess multiple digital identities issued by different identity services, as depicted in Figure~\ref{fig:access-control}.
Each identity, including its certificate $cert$ and private key $prikey$, is stored in a blockchain wallet $W$.
Therefore, the platform allows an IoT device to register its service in multiple organizations using different identities stored in the same wallet, i.e., $W = \{(cert_i, prikey_i)| i=1,2,...,N\}$.

The platform enforces more granular access control via a multi-layer RBAC model, shown in Figure~\ref{fig:access-control}.
In this model, the digital identity of each actor is assigned a role when it is created.
The first layer, the blockchain-level access control policies, define which organizations and roles can query or update the ledger using smart contracts.
By default, actors of a participating organization can see all registered IoT devices and services, but only actors with the ``writer'' roles have the right to register or update them.
On the next layer, at the transaction level, service registry and service broker smart contracts can be attached with transaction validation policies.
These policies tell which organizations and actors must sign the service request transaction for it to be valid.
Finally, the smart contracts enforce the last layer of access control for each IoT device, service, service request, and service response.
The smart contracts check the caller's identity to ensure that only the service-owning IoT device can update the device and service information and respond to service requests.

\subsection{Service Request \& Response}

\begin{table*}[htb]
\caption{Service request and response information.\label{tbl:service-request-response-information}}
\begin{tabularx}{\textwidth}{m{1.25in}m{1.25in}l}
\toprule
Category                 & Field & Description                                          \\
\midrule
\multirow{5}{*}{Service Request}  & ID             & UUID of the request.                                          \\
                                  & Time           & Date and time when the request is created.                    \\
                                  & Service        & Name, device ID, and organization ID of the requested service.\\
                                  & Method         & Request method.                                               \\
                                  & Arguments      & Optional arguments of the request.                            \\
\midrule
\multirow{4}{*}{Service Response} & Request ID     & UUID of the request to which the response answers.            \\
                                  & Time           & Date and time when the response is created.                   \\
                                  & Status Code    & The request processing status.                                \\
                                  & Return Value   & Optional value to be returned to the service request sender.  \\
\bottomrule
\end{tabularx}
\end{table*}

\begin{figure}[t]
\centering
\includegraphics[width=\linewidth]{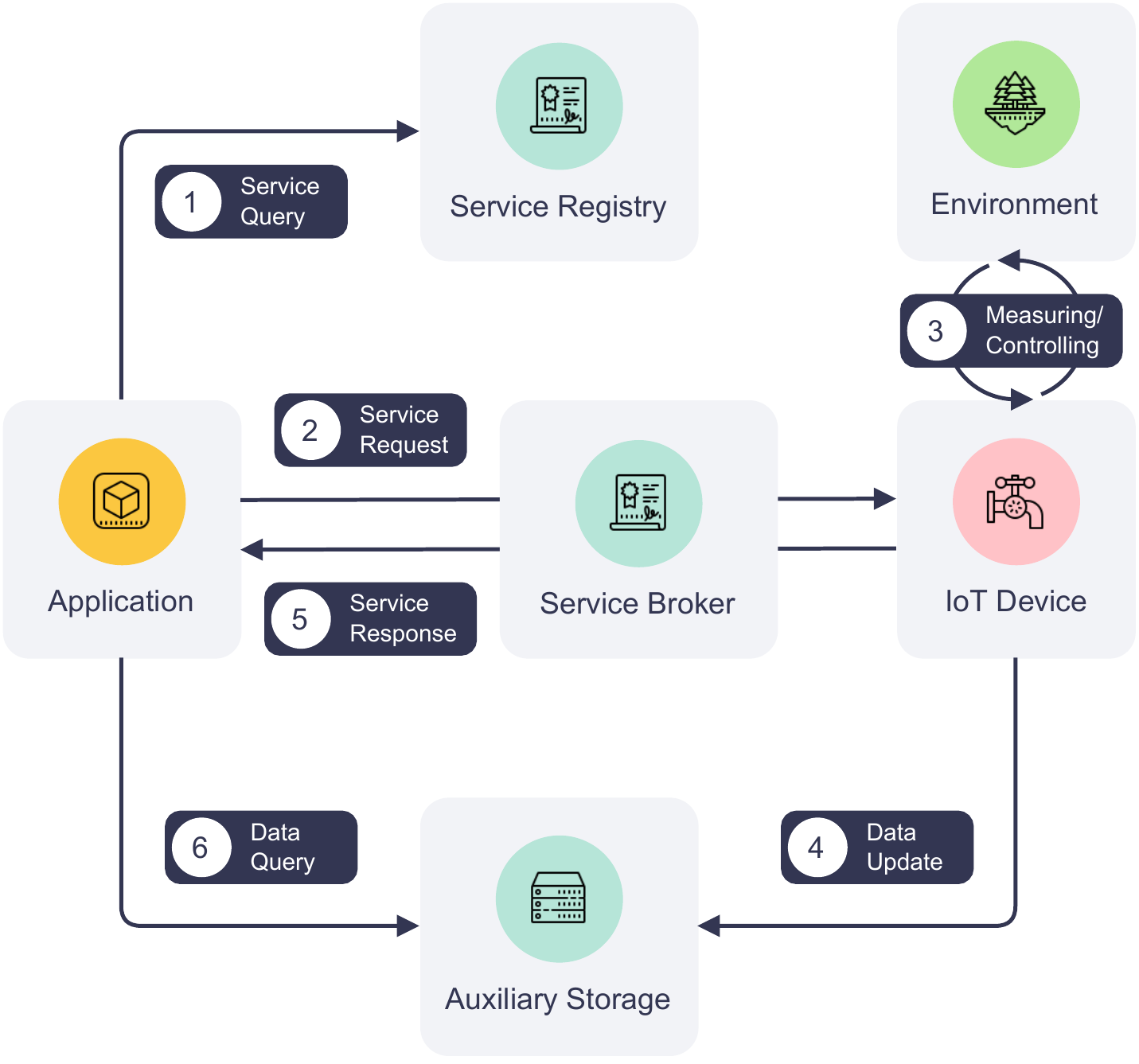}
\caption{Process of communication between an IoT device and application.%
\label{fig:communication-process}}
\end{figure}

The consortium blockchain provides opportunities for IoT networks to become decentralized and distributed.
It also enables sharing of IoT infrastructure among organizations within a consortium in a scalable and reliable way.
Given the advantages of consortium blockchain, we propose an IoT communication process that utilizes the blockchain network as the communication channel.
In this process, data exchange between parties is achieved in service requests and responses facilitated by the service broker smart contract.
Figure~\ref{fig:communication-process} provides a more detailed look into step \circled{3} and \circled{4} of Figure~\ref{fig:service-lifecycle}.
It illustrates the communication process between an IoT device and an application, including querying available services, requesting services, responding to service requests, and retrieving data.

1) \textit{Service querying}:
To find out a service's availability, an application can query the service registry before initiating a service request, as illustrated by step \circled{1} in Figure~\ref{fig:communication-process}.
Given the service name, device ID, and organization, the service registry returns information about the service to the application, such as the service version and last update time.
Although this step is not mandatory, an application is recommended to refresh service information periodically to stay updated on the service's information and look for alternative services when the service is unavailable.
Additionally, the last update time of service helps decide the best time to refresh such information.

2) \textit{Service request and processing}:
To communicate with an IoT device on the proposed platform, an application creates a service request first.
The fields of a service request are listed in Table~\ref{tbl:service-request-response-information}.
Every service request is identified by a universally unique identifier (UUID) generated by the application.
The request ID must be unique among all requests on the platform.
Information about the requested service, such as the service name, is required so that the service-providing IoT device can retrieve the request from the blockchain.
The request also contains the request creation time used for request deduplication and auditing purposes.
The request body is represented by the method and arguments fields where the action and its optional arguments are defined.

Next, the application submits the service request to the service broker, which will validate the request and create a blockchain transaction.
Once the transaction is validated and endorsed by the network and written to the ledger, the service-providing IoT device will be notified by the network that its service has been requested.
An IoT device can then fetch request information from the blockchain via the service broker and interact with its environment as instructed by the request.
These processes are depicted in step \circled{2} and \circled{3} in Figure~\ref{fig:communication-process}.

3) \textit{Service response and data retrieval}:
IoT devices can respond to service requests asynchronously whenever they finish processing them.
The process of replying to a service request is shown in step \circled{4} to \circled{6} in Figure~\ref{fig:communication-process}.
An IoT device starts by creating the service response containing the information described in the bottom half of Table~\ref{tbl:service-request-response-information}.
Apart from the UUID of the corresponding request, the response also contains fields about response creation time, a custom status code indicating the status of processing, and an optional value to be returned to the requester.
The device may also store operational data in the auxiliary storage and leave a pointer to the data in the return value under situations where the volume or time sensitivity of the data cannot be met by the blockchain.
Details of the auxiliary storage are further discussed in Section~\ref{sec:data-storage}.

Following its creation, the service response is sent by the IoT device to the service broker, which creates another blockchain transaction.
The transaction will undergo the same process of validation and endorsement as the service request transaction and eventually be appended to the distributed ledger.
By listening to transaction events on the blockchain, an application can act upon the completion of its request, e.g., retrieving retrieve the response from the blockchain.
Furthermore, if the return value contains a pointer to data in the auxiliary storage, the application can perform additional storage operations to retrieve the data.

\subsection{Data Storage}
\label{sec:data-storage}

\begin{figure}[b]
\centering
\includegraphics[width=\linewidth]{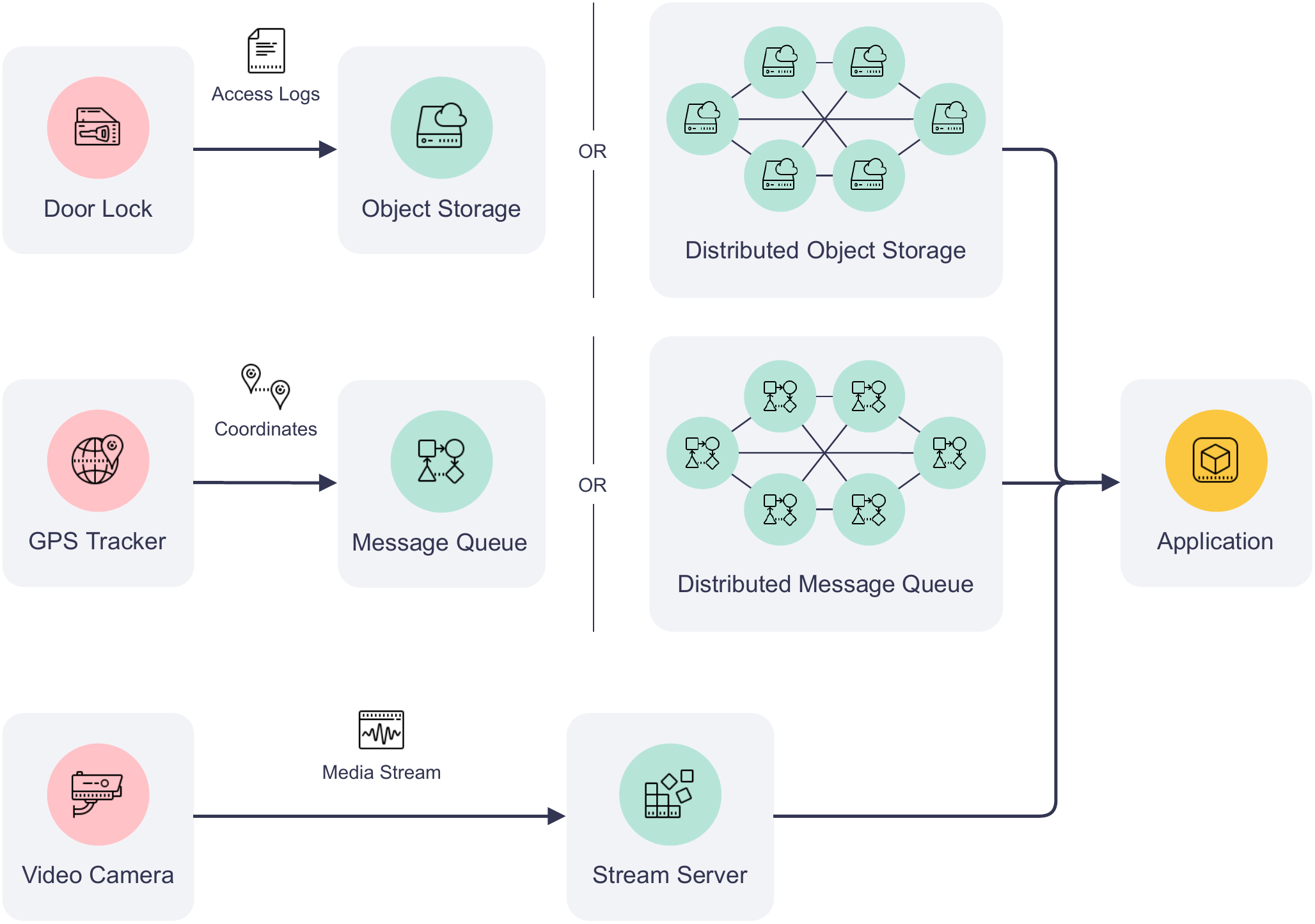}
\caption{Auxiliary storage types and applications.\label{fig:auxiliary_storage}}
\end{figure}

Due to blockchain's transaction size, throughput, and latency limitations, oftentimes it is preferable to store IoT-generated data off-chain.
To meet the storage requirements of different IoT applications, we propose auxiliary storage to complement the data exchange on the platform.
Figure~\ref{fig:auxiliary_storage} demonstrates how auxiliary storage facilitates IoT devices to pass the data to an application.
The storage system can be composed of different types of data storage depending on the communication requirements.
For example, non-real-time structured data can be stored in the cloud such as AWS S3, or a distributed object storage like the InterPlanetary File System (IPFS).
IoT device logs and historical sensor readings fall into this category.
Meanwhile, real-time sensor data (e.g., instantaneous geolocation coordinates) can be published to message queues such as Apache Kafka for efficient data delivery.
Finally, media streaming servers can be included in the auxiliary storage to transcode, cache, and stream multimedia captured by IoT devices to their users.
Examples of this data type are surveillance camera feed and audio data collected by an acoustic gunshot detection system.
Each type of storage can be offered by a single organization of the consortium or hosted by multiple organizations collaboratively.
The design of data storage is out of the scope of this paper.

\section{Implementation and Case Studies}
\label{sec:implementation}

This section dives into the implementation details of the proposed platform.
We further showcase the capabilities and usefulness of the platform with two real-world IoT applications.
For the sake of research reproducibility, the source code of our implementation, the exemplary case study applications, and all scripts for testbed setup and benchmarking are made available online\footnote{Please refer to \hyperlink{sec:data-availability-statement}{Data Availability Statement}}.

\subsection{Platform Implementation}

Considering the feature richness and development support, we selected Hyperledger Fabric \cite{Androulaki2018} as the consortium blockchain platform for our IoT service platform implementation.
Thanks to its modular architecture and plug-and-play nature, Hyperledger Fabric has been widely used in industrial environments.
Our implementation takes advantage of Hyperledger Fabric's components to realize the core functionalities of the proposed platform in the following aspects:

1) We incorporated device and user identities into the blockchain using Hyperledger Fabric's membership service provider (MSP).
Each organization on the platform has its dedicated MSP, which translates the identities into roles and privileges of the blockchain.
Thus, the platform can authenticate invocations to smart contracts using registered identities.

2) Hyperledger Fabric defines various policies agreed by the consortium members, or channel members in Hyperledger Fabric's terminology, for
infrastructure management.
In our implementation, we limit IoT devices and users to only submitting transactions or querying the ledger using ACL.
We also restrict which organizations must approve or endorse the transactions with smart contract endorsement policies.
In addition, smart contracts also limit access to write operations like responding to service requests to the device and service owners by checking the caller's identity.

3) The smart contracts of the proposed platform were implemented using Fabric contract API in the Go programming language.
All smart contracts are packaged in one chaincode, a container for smart contracts, so that they share the same world state.
In addition, we created SDK for Go, Java, JavaScript, and TypeScript programming languages to simplify application development for our proposed platform.

4) Client communications with the blockchain are simplified using the Hyperledger Fabric gateway.
Instead of directly interacting with the blockchain network and managing the complexities of transaction proposal, endorsement, and commission, IoT devices and users now delegate most of the heavy lifting operations to a gateway component running on peer nodes.
This improvement is essential for devices that are energy-constrained or low on computing power.

\subsection{Case Study: Parrot}

\begin{figure}[htb]
\centering
\includegraphics[width=\linewidth]{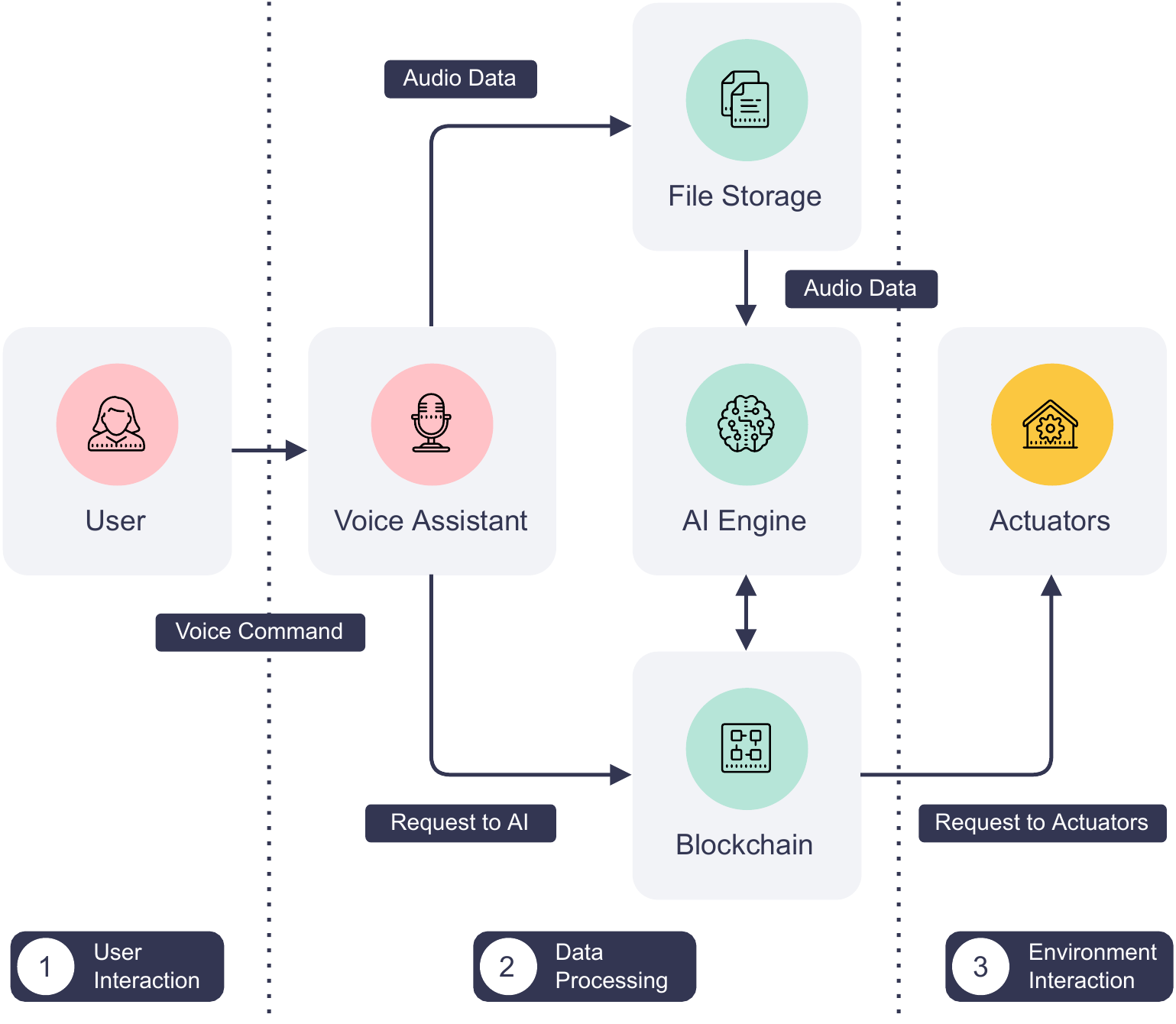}
\caption{The architecture and data flow of Parrot.\label{fig:parrot-arch}}
\end{figure}

Parrot is a voice assistant for the smart home lighting system implemented using the proposed IoT service blockchain.
It enables touch-less control of home lights using voice commands like "Parrot, turn on the kitchen light."
The architecture of Parrot is shown in Figure~\ref{fig:parrot-arch}.
The workflow starts with a user speaking the voice command of a given format to the smart speaker.
A microphone on the smart speaker continuously listens in the background but only begins recording voice commands on wake words, or "Parrot" in this case.
It also determines the duration of the recording and stops when the command ends.
This is done using an onboard wake word detection engine, Porcupine \cite{porcupine}, and the voice activity detector provided by WebRTC \cite{webrtc_vad}.
Then, the smart speaker adds the recording to a private decentralized file storage network, implemented by IPFS, and calls the service exposed by a remote voice AI engine to the IoT service platform.
Next, the AI engine is notified by the blockchain and retrieves the user's audio recording from IPFS.
The data is then fed into Rhino \cite{rhino}, a speech-to-intent engine that decodes voice commands and extracts the location of the light and actions to perform.
Finally, the AI engine sends requests for turning on or off to the corresponding actuator, or smart light, via the blockchain.

The main advantage of this model is that all communications are securely backed by the proposed platform.
Compared to traditional IoT systems, actors of this system publish their services only to the blockchain instead of exposing them using other insecure communication protocols.
Moreover, all processes involved are transparent to the user since service activities are recorded on the immutable blockchain.
Blockchain transactions leave an audit trail that is invaluable to incident response and forensic investigation when a problem arises.
Regarding user privacy, the network operator can isolate different user groups using separate Hyperledger Fabric channels.
Audio recordings of the user can also be set to expire automatically by unpinning and garbage collecting them from the storage network.

\subsection{Case Study: Crystal Ball}

Surveillance cameras are widely deployed nowadays.
However, they are often vulnerable to hackers or malware due to poor security design and improper configuration.
An unprotected surveillance camera can seriously threaten user privacy as the video footage may be leaked to unauthorized users.
For example, Insecam \footnote{http://www.insecam.org} is a live camera directory that allows visitors to view live streams from thousands of unprotected public cameras as of May 2022, and the number of exposed cameras is still growing.
Hackers can also use a compromised camera in other cybercrimes, e.g., to form a botnet and initiate distributed denial-of-service (DDoS) attacks.

\begin{figure}[htb]
\centering
\includegraphics[width=\linewidth]{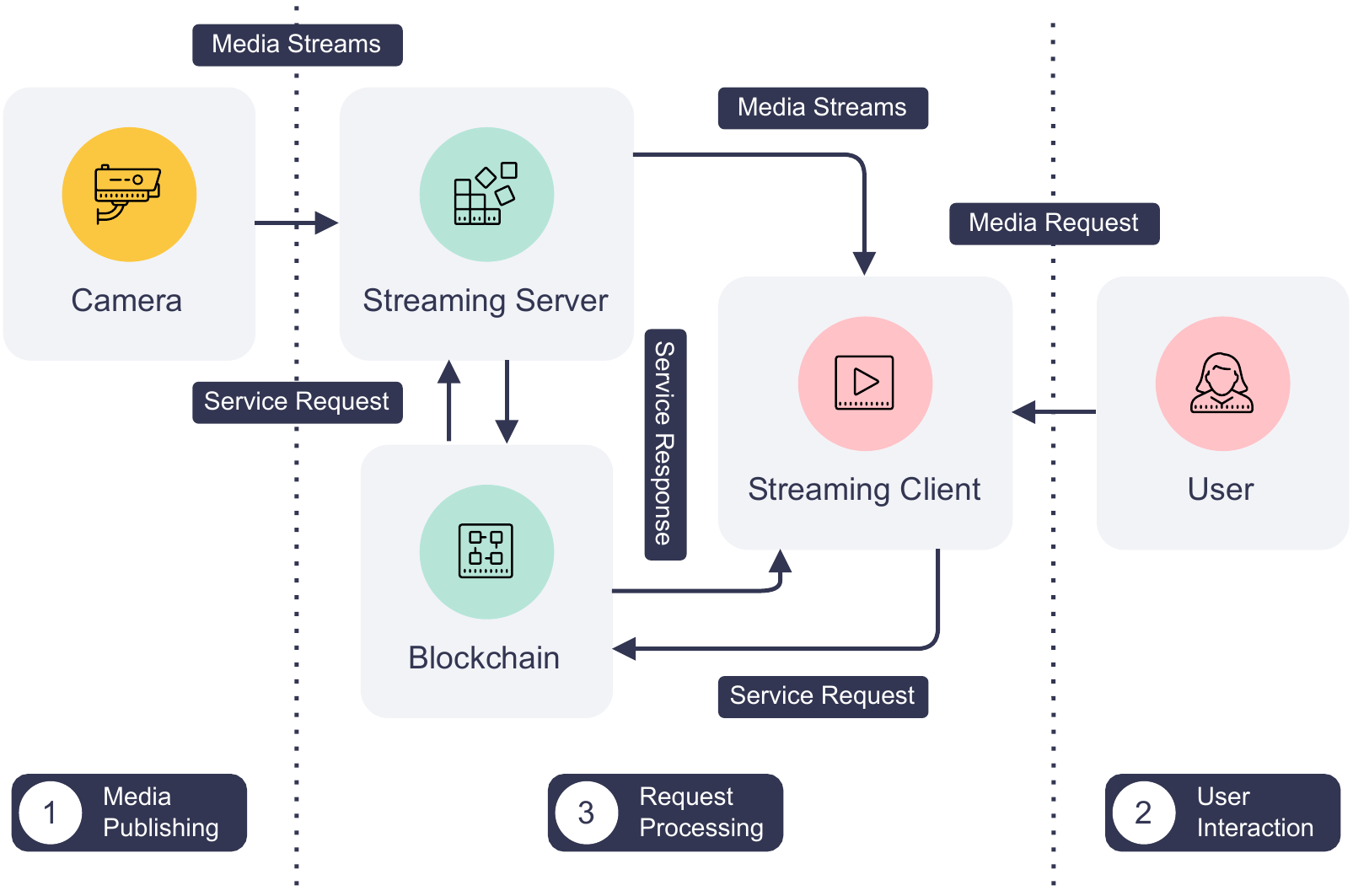}
\caption{The architecture and data flow of Crystal Ball.\label{fig:crystal-ball-arch}}
\end{figure}

We have built a blockchain-based secure surveillance streaming system called Crystal Ball using the proposed IoT service platform.
Figure~\ref{fig:crystal-ball-arch} depicts the architecture and workflow of Crystal Ball.
A camera in the Crystal Ball system does not serve its video and audio feeds on an open network port.
Instead, it publishes them to a secure streaming server that streams camera feeds only to users with correct access tokens.
Meanwhile, the streaming server creates an IoT service that generates and distributes one-time session access tokens to authorized users on the blockchain.
Finally, users can request access tokens and watch live streams from an Android streaming client.

Crystal Ball eliminates the need for exposing camera feeds from the capturing device.
Unnecessary ports can now be closed to reduce the attack surface of the camera.
Furthermore, Crystal Ball protects camera feeds from unauthorized access using device identities and access tokens.
As access tokens are one-time only and tied to each session, a malicious user cannot reuse previous tokens even if they are leaked.
Lastly, an administrator can easily log and analyze camera accesses using transaction history and decommission cameras that are compromised by invalidating their device identities.

\section{Evaluation and Discussion}
\label{sec:evaluation}

In this section, we present the experiment settings under which we evaluate the proposed IoT service platform.
Our evaluation focuses on two core configuration parameters of the blockchain, transaction batch size and batch timeout, to discover the optimal blockchain configuration for our platform.
We also explore the transaction performance overhead introduced by additional IoT device connections.
Transaction throughput, transaction latency, and system resource utilization are measured in each test as the performance indicators of the platform.
Additionally, we examine the proposed solution regarding its security and privacy impact on IoT systems.

\subsection{Experimental Setup and Methodology}

\begin{figure}[b]
\centering
\includegraphics[width=\linewidth]{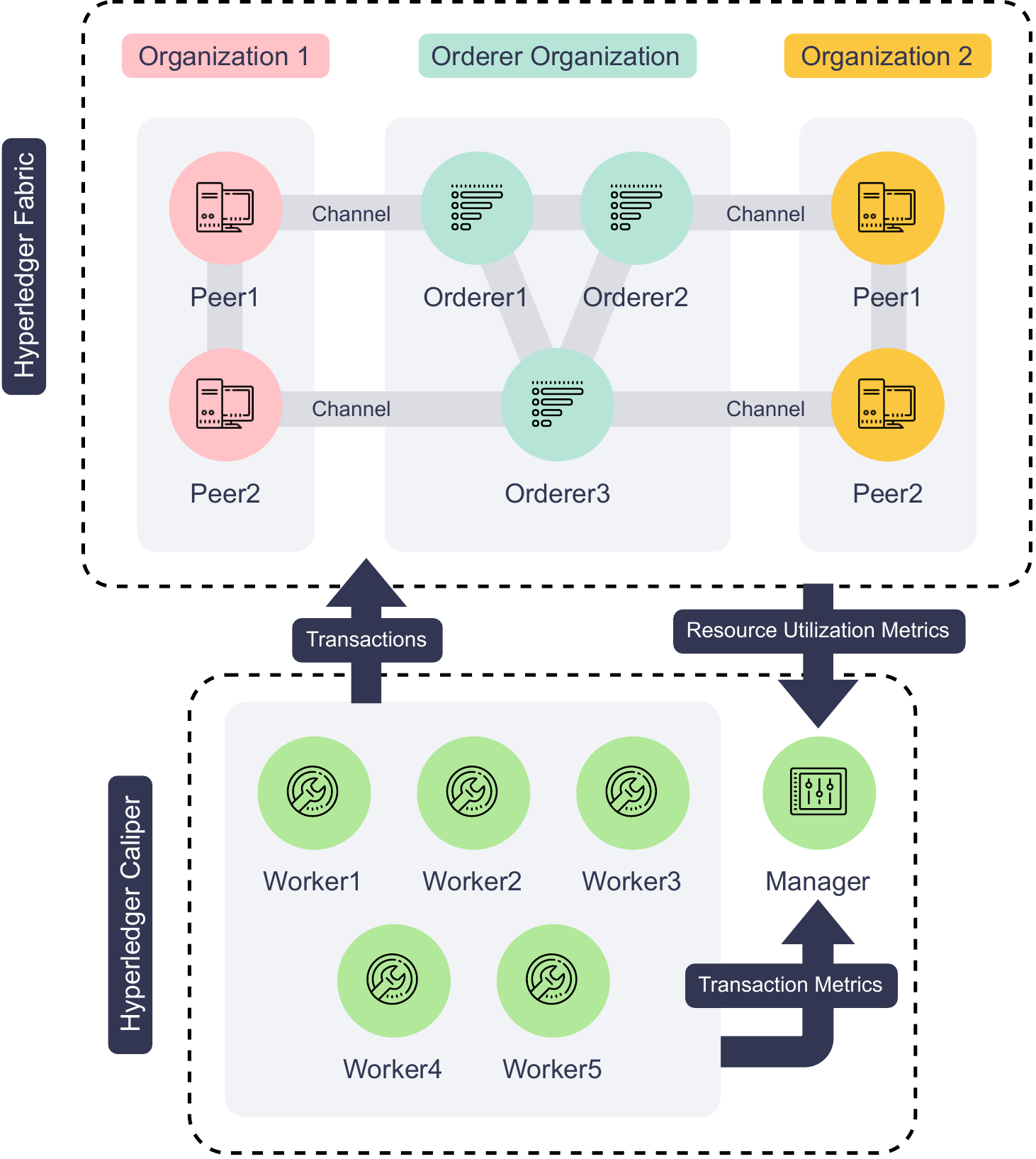}
\caption{IoT service platform testbed architecture.\label{fig:testbed-arch}}
\end{figure}

Figure~\ref{fig:testbed-arch} illustrates the architecture of our testbed and Table~\ref{tbl:testbed-hardware} details the hardware and software configuration of every machine used in our experiments.
The proposed IoT service platform runs on a multi-node Hyperledger Fabric blockchain comprising two peer organizations and one orderer organization.
Each peer organization contains two peer nodes; the order organization contains three Raft orderer nodes, and all are running Hyperledger Fabric version v2.4.3.
All organizations form a single consortium, and transactions are executed and ordered in a single channel.
The channel enforces the "MAJORITY" endorsement policy, which means two peer organizations must all endorse the transaction for it to be committed to the blockchain.
Then, the IoT service blockchain chaincode is installed on all peer nodes.
To resemble a multi-organizational environment, we have deployed the testing blockchain on bare-metal servers across two Chameleon Cloud \cite{Keahey2019} data centers.
Compared to deploying all nodes to the same data center, our set-up will introduce a network latency of around 30ms between all consortium parties to reflect the real-world network environment of a blockchain network.

Apart from the Hyperledger Fabric blockchain nodes, six virtual machines were employed to establish the Hyperledger Caliper \cite{Caliper} benchmarking environment.
Hyperledger Caliper is a blockchain benchmarking tool that generates synthetic transaction workloads and measures the performance of the system under test (SUT).
Our benchmarking environment is composed of a Caliper manager and five Caliper workers.
The manager distributes workload parameters to the workers, synchronizes workers during each test, collects evaluation results, and generates human-readable reports.
Each of the five workers starts two client connections to the blockchain to simulate ten IoT devices that provide or request services on the
proposed platform.
These workers execute the workload scripts to call the smart contracts and generate transactions guided by the \lstinline{fixed-load} rate control strategy.
This strategy means the workers send the transactions or queries at a dynamic rate such that the number of incomplete transactions or queries in the SUT always stays under a given value.
In our experiments, the workers collectively send 2,000 transactions or queries under a fixed load of 100 transactions/queries.

\begin{table*}[tbp]
\caption{Hardware configurations of Hyperledger Fabric nodes and
Hyperledger Caliper nodes.\label{tbl:testbed-hardware}}
\begin{tabularx}{\textwidth}{lrrrrl}
\toprule
Node Name            & CPU (Cores) & Memory (GB) & Disk (GB) & Network (Gbps) & Location      \\
\midrule
Hyperledger Fabric Orderer1   & 24                   & 128                  & 233                & 1                       & Chicago, IL, USA       \\
Hyperledger Fabric Orderer2   & 24                   & 128                  & 233                & 1                       & Chicago, IL, USA       \\
Hyperledger Fabric Orderer3   & 24                   & 128                  & 233                & 1                       & Chicago, IL, USA       \\
Hyperledger Fabric Org1 Peer1 & 48                   & 191                  & 447                & 10                      & Austin, TX, USA        \\
Hyperledger Fabric Org1 Peer2 & 48                   & 191                  & 447                & 10                      & Austin, TX, USA        \\
Hyperledger Fabric Org2 Peer1 & 48                   & 191                  & 447                & 10                      & Austin, TX, USA        \\
Hyperledger Fabric Org2 Peer2 & 48                   & 191                  & 447                & 10                      & Austin, TX, USA        \\
Hyperledger Caliper Manager   & 2                    & 2                    & 40                 & 1                       & Chattanooga, TN, USA   \\
Hyperledger Caliper Worker1   & 2                    & 2                    & 40                 & 1                       & Chattanooga, TN, USA   \\
Hyperledger Caliper Worker2   & 2                    & 2                    & 40                 & 1                       & Chattanooga, TN, USA   \\
Hyperledger Caliper Worker3   & 2                    & 2                    & 40                 & 1                       & Chattanooga, TN, USA   \\
Hyperledger Caliper Worker4   & 2                    & 2                    & 40                 & 1                       & Chattanooga, TN, USA   \\
Hyperledger Caliper Worker5   & 2                    & 2                    & 40                 & 1                       & Chattanooga, TN, USA   \\
\bottomrule
\end{tabularx}
\end{table*}

The core metrics inspected in each test are peak read/transaction throughput and average read/transaction latency.
According to the Hyperledger Blockchain Performance Metrics white paper \cite{HyperledgerMetrics2018}, a read operation does not change the ledger state, while a transaction operation involves ledger updates.
The latency $L_{read}$ of a read operation measured in seconds is defined as:
\begin{equation}
L_{read} = t_{response} - t_{submit},
\end{equation}
where $t_{submit}$ is the read request submission time and $t_{response}$ is the time at which the reply is received.
The throughput $W_{read}$ of read operations measured in read per second (RPS) is defined as:
\begin{equation}
W_{read} = \frac{N_{read}}{T_{read}},
\end{equation}
where $N_{read}$ is the total number of read operations completed in time $T_{read}$.

Transaction throughput and latency are measured differently from read throughput and latency since the confirmation time of blocks must be considered in transaction operations.
The transaction latency $L_{tx}$ in seconds is defined as:
\begin{equation}
L_{tx} = t_{confirm} - t_{submit},
\end{equation}
where $t_{confirm}$ is the time at which the transaction is confirmed by the network given a network threshold (e.g., 90\% of the network) and $t_{submit}$ is the time transaction is submitted by the client.
The transaction throughput $W_{tx}$ measured in transaction per second (TPS) is defined as:
\begin{equation}
W_{tx} = \frac{N_{tx}}{T_{tx}},
\end{equation}
where $N_{tx}$ is the total number of committed transactions at all nodes of the network in time $T_{tx}$.

System resource utilization metrics are also measured during each test.
These include average CPU and memory usage, total data sent to or received from the network, and total data read from or written to the disk.
These metrics are polled and aggregated by Prometheus \cite{prometheus_2022} periodically from each Hyperledger Fabric node and reported to the Caliper manager.
The read/transaction throughput and latency of the blockchain and the above resource utilization metrics indicate how well our proposed platform performs under different blockchain configurations.
They also provide valuable information on optimizing the blockchain for various use cases.

\subsection{Performance vs. Batch Size}

We first examine the impact of transaction batch size on the performance of the proposed platform.
Three parameters constrain the batch size in the Hyperledger Fabric blockchain: maximum message count $N_{max}$, preferred maximum bytes of messages $S_{soft}$, and absolute maximum bytes of messages $S_{hard}$.
Therefore, the batch size configuration directly controls the number of messages $N_{msg}$ that can be batched into a block.
When the size of each message, or transaction data, is relatively small, $N_{msg}$ is primarily limited by $N_{max}$.
Otherwise, $N_{msg}$ will mostly be limited by $S_{soft}$ unless there are many huge messages whose sizes exceed $S_{hard}$.

In our experiments, we intend to control the batch size with $S_{soft}$ solely, so we assign large constant values to $N_{max}$ and $S_{hard}$ to ensure blocks always reach $S_{soft}$ first.
Then, we set $S_{soft}$ to 128KB, 256KB, 512KB, 1MB, 2MB, 4MB, and 8MB and evaluate the smart contract operations of the proposed platform.
Device registration and deregistration transaction throughput and latency are omitted from the results since the workers only start ten blockchain clients for the tests.
As each IoT device binds to one client identity, the workers are limited to registering or deregistering ten devices in total.
Therefore, few transactions can be evaluated for the above two operations to report their accurate throughput and latency results, and we exclude these operations from the result analysis.
However, we will revisit these operations and their performance in later experiments.

\begin{figure}[b]
\centering
\begin{tikzpicture}
    \begin{groupplot}[
        group style={group size=1 by 2},
        xmin=0,
        ymin=0,
        width=.9\linewidth,
        height=1.75in,
        xlabel={Batch Size (MB)},
        xtick distance=1,
    ]
    \nextgroupplot[ylabel={Throughput (RPS)}]
        \addplot+ table [x={Batch Size (MB)}, y={/device-registry/get}, col sep=tab] {Plots/throughput-vs-batch-size.csv};
        \addplot+ table [x={Batch Size (MB)}, y={/device-registry/get-all}, col sep=tab] {Plots/throughput-vs-batch-size.csv};
        \addplot+ table [x={Batch Size (MB)}, y={/service-registry/get}, col sep=tab] {Plots/throughput-vs-batch-size.csv};
        \addplot+ table [x={Batch Size (MB)}, y={/service-registry/get-all}, col sep=tab] {Plots/throughput-vs-batch-size.csv};
        \addplot+ table [x={Batch Size (MB)}, y={/service-broker/get}, col sep=tab] {Plots/throughput-vs-batch-size.csv};
        \addplot+ table [x={Batch Size (MB)}, y={/service-broker/get-all}, col sep=tab] {Plots/throughput-vs-batch-size.csv};
    
    \nextgroupplot[ylabel={Latency (s)}, ytick distance=3]
        \addplot+ table [x={Batch Size (MB)}, y={/device-registry/get}, col sep=tab] {Plots/latency-vs-batch-size.csv};
        \addplot+ table [x={Batch Size (MB)}, y={/device-registry/get-all}, col sep=tab] {Plots/latency-vs-batch-size.csv};
        \addplot+ table [x={Batch Size (MB)}, y={/service-registry/get}, col sep=tab] {Plots/latency-vs-batch-size.csv};
        \addplot+ table [x={Batch Size (MB)}, y={/service-registry/get-all}, col sep=tab] {Plots/latency-vs-batch-size.csv};
        \addplot+ table [x={Batch Size (MB)}, y={/service-broker/get}, col sep=tab] {Plots/latency-vs-batch-size.csv};
        \addplot+ table [x={Batch Size (MB)}, y={/service-broker/get-all}, col sep=tab] {Plots/latency-vs-batch-size.csv};
    \end{groupplot}
\end{tikzpicture}

\vspace{0.25cm}
\begin{tikzpicture}
    \matrix [fill=white,draw=black,cells={anchor=west}]{
        \LegendImage{linestyle1} & \LegendEntry{Query a device}; &
        \LegendImage{linestyle2} & \LegendEntry{Query all devices}; &
        \LegendImage{linestyle3} & \LegendEntry{Query a service}; \\
        \LegendImage{linestyle4} & \LegendEntry{Query all services}; &
        \LegendImage{linestyle5} & \LegendEntry{Query a request}; &
        \LegendImage{linestyle6} & \LegendEntry{Query all requests}; \\
    };
\end{tikzpicture}
\caption{Throughput and latency of read operations for varying batch sizes.\label{fig:performance-read-vs-batch-size}}
\end{figure}
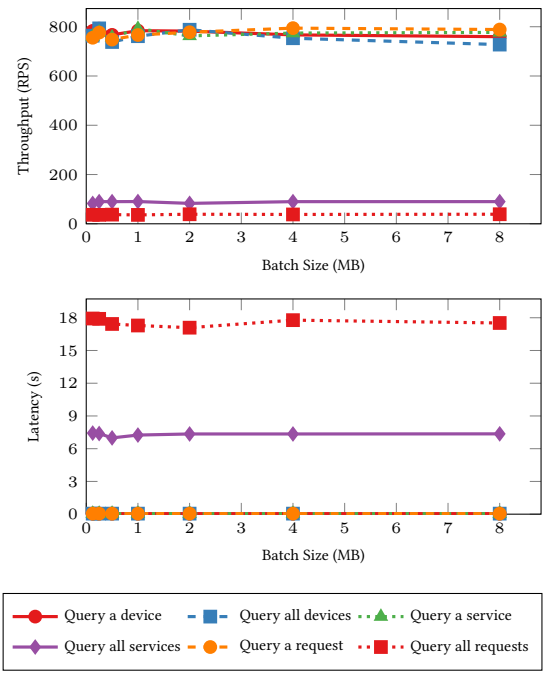

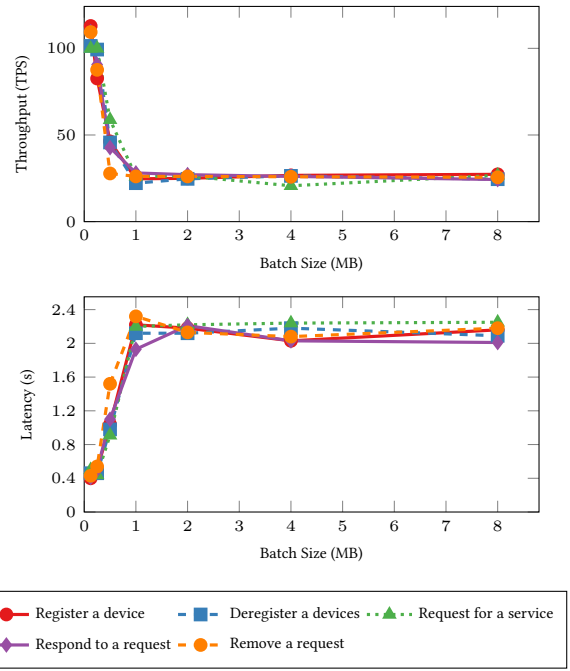
\begin{figure}[t]
\centering
\begin{tikzpicture}
    \begin{groupplot}[
        group style={group size=1 by 2},
        xmin=0,
        ymin=0,
        width=.9\linewidth,
        height=1.75in,
        xlabel={Batch Size (MB)},
        xtick distance=1,
    ]
    \nextgroupplot[ylabel={Throughput (TPS)}]
        \addplot+ table [x={Batch Size (MB)}, y={/service-registry/register}, col sep=tab] {Plots/throughput-vs-batch-size.csv};
        \addplot+ table [x={Batch Size (MB)}, y={/service-registry/deregister}, col sep=tab] {Plots/throughput-vs-batch-size.csv};
        \addplot+ table [x={Batch Size (MB)}, y={/service-broker/request}, col sep=tab] {Plots/throughput-vs-batch-size.csv};
        \addplot+ table [x={Batch Size (MB)}, y={/service-broker/respond}, col sep=tab] {Plots/throughput-vs-batch-size.csv};
        \addplot+ table [x={Batch Size (MB)}, y={/service-broker/remove}, col sep=tab] {Plots/throughput-vs-batch-size.csv};
    
    \nextgroupplot[ylabel={Latency (s)}, ytick distance=0.4]
        \addplot+ table [x={Batch Size (MB)}, y={/service-registry/register}, col sep=tab] {Plots/latency-vs-batch-size.csv};
        \addplot+ table [x={Batch Size (MB)}, y={/service-registry/deregister}, col sep=tab] {Plots/latency-vs-batch-size.csv};
        \addplot+ table [x={Batch Size (MB)}, y={/service-broker/request}, col sep=tab] {Plots/latency-vs-batch-size.csv};
        \addplot+ table [x={Batch Size (MB)}, y={/service-broker/respond}, col sep=tab] {Plots/latency-vs-batch-size.csv};
        \addplot+ table [x={Batch Size (MB)}, y={/service-broker/remove}, col sep=tab] {Plots/latency-vs-batch-size.csv};
    \end{groupplot}
\end{tikzpicture}

\vspace{0.25cm}
\begin{tikzpicture}
    \matrix [fill=white,draw=black,cells={anchor=west}]{
        \LegendImage{linestyle1} & \LegendEntry{Register a device}; &
        \LegendImage{linestyle2} & \LegendEntry{Deregister a devices}; &
        \LegendImage{linestyle3} & \LegendEntry{Request for a service}; \\
        \LegendImage{linestyle4} & \LegendEntry{Respond to a request}; &
        \LegendImage{linestyle5} & \LegendEntry{Remove a request}; \\
    };
\end{tikzpicture}
\caption{Throughput and latency of transaction operations for varying batch sizes.\label{fig:performance-transaction-vs-batch-size}}
\end{figure}

Figure~\ref{fig:performance-read-vs-batch-size} and Figure~\ref{fig:performance-transaction-vs-batch-size} present the throughput and latency of read and transaction operations under various batch sizes.
The throughput and latency of all read operations stay nearly unchanged as batch size increases.
This observation is because read operations in Hyperledger Fabric are not sent to the ordering service for validation and committing to the ledger \cite{baliga_performance_2018}.
Thus, batch size configuration has no impact on the performance of read operations.
On the other hand, the size of messages has more impact on the performance metrics.
The ``query all requests'' and ``query all services'' operations have lower throughput and higher latency as the messages returned to the clients are significantly larger than the singular queries (e.g., ``query a device'') and device queries.
It can take more time and bandwidth for the peers to process and transmit such messages, resulting in lower performance.
In contrast, the performance metrics of transaction operations are tied more closely to batch sizes.
Throughput and latency degrade quickly from 100TPS and 0.4s to 25TPS and 2.4s as batch size increases until it reaches 1MB since larger blocks need more time to generate and dispatch.
Further increasing the batch size will not worsen the performance.

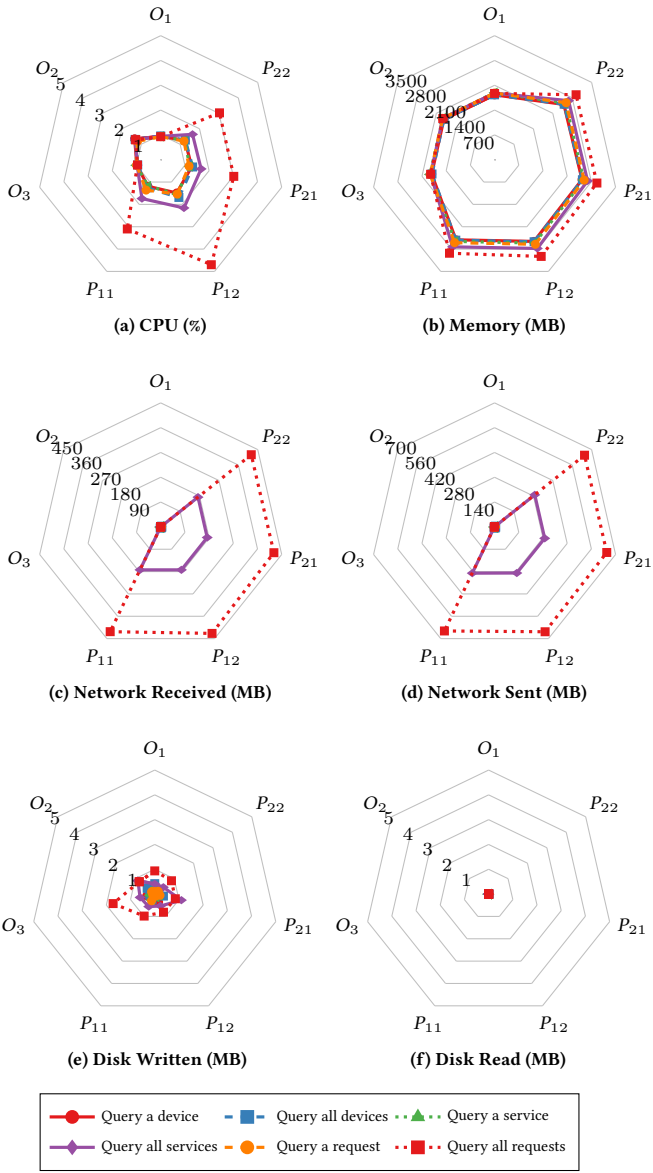
\begin{figure}[htb]
\centering
\captionsetup[subfigure]{justification=centering}
\subfloat[CPU (\%)]{
    \begin{tikzpicture}[font=\footnotesize]
        \tkzKiviatDiagram[
            gap=.6,
            scale=.55,
            rotate=90,
            lattice=5,
            label space=.5,
            label style/.style={text width=1em,align=center},
            radial style/.style={draw opacity=0},
        ]{
            $O_1$,
            $O_2$,
            $O_3$,
            $P_{11}$,
            $P_{12}$,
            $P_{21}$,
            $P_{22}$
        }
        \tkzKiviatLineFromFile[linestyle1]{Plots/2mbx2s-cpu.csv}{1}
        \tkzKiviatLineFromFile[linestyle2]{Plots/2mbx2s-cpu.csv}{2}
        \tkzKiviatLineFromFile[linestyle3]{Plots/2mbx2s-cpu.csv}{4}
        \tkzKiviatLineFromFile[linestyle4]{Plots/2mbx2s-cpu.csv}{5}
        \tkzKiviatLineFromFile[linestyle5]{Plots/2mbx2s-cpu.csv}{9}
        \tkzKiviatLineFromFile[linestyle6]{Plots/2mbx2s-cpu.csv}{10}
        \tkzKiviatGrad[graduation distance=-13,unity=1](1)
    \end{tikzpicture}
}
\subfloat[Memory (MB)]{
    \begin{tikzpicture}[font=\footnotesize]
        \tkzKiviatDiagram[
            gap=.6,
            scale=.55,
            rotate=90,
            lattice=5,
            label space=.5,
            label style/.style={text width=1em,align=center},
            radial style/.style={draw opacity=0},
        ]{
            $O_1$,
            $O_2$,
            $O_3$,
            $P_{11}$,
            $P_{12}$,
            $P_{21}$,
            $P_{22}$
        }
        \tkzKiviatLineFromFile[linestyle1]{Plots/2mbx2s-memory.csv}{1}
        \tkzKiviatLineFromFile[linestyle2]{Plots/2mbx2s-memory.csv}{2}
        \tkzKiviatLineFromFile[linestyle3]{Plots/2mbx2s-memory.csv}{4}
        \tkzKiviatLineFromFile[linestyle4]{Plots/2mbx2s-memory.csv}{5}
        \tkzKiviatLineFromFile[linestyle5]{Plots/2mbx2s-memory.csv}{9}
        \tkzKiviatLineFromFile[linestyle6]{Plots/2mbx2s-memory.csv}{10}
        \tkzKiviatGrad[graduation distance=-15,unity=700](1)
    \end{tikzpicture}
}\\
\subfloat[Network Received (MB)]{
    \begin{tikzpicture}[font=\footnotesize]
        \tkzKiviatDiagram[
            gap=.6,
            scale=.55,
            rotate=90,
            lattice=5,
            label space=.5,
            label style/.style={text width=1em,align=center},
            radial style/.style={draw opacity=0},
        ]{
            $O_1$,
            $O_2$,
            $O_3$,
            $P_{11}$,
            $P_{12}$,
            $P_{21}$,
            $P_{22}$
        }
        \tkzKiviatLineFromFile[linestyle1]{Plots/2mbx2s-network-in.csv}{1}
        \tkzKiviatLineFromFile[linestyle2]{Plots/2mbx2s-network-in.csv}{2}
        \tkzKiviatLineFromFile[linestyle3]{Plots/2mbx2s-network-in.csv}{4}
        \tkzKiviatLineFromFile[linestyle4]{Plots/2mbx2s-network-in.csv}{5}
        \tkzKiviatLineFromFile[linestyle5]{Plots/2mbx2s-network-in.csv}{9}
        \tkzKiviatLineFromFile[linestyle6]{Plots/2mbx2s-network-in.csv}{10}
        \tkzKiviatGrad[graduation distance=-15,unity=90](1)
    \end{tikzpicture}
}
\subfloat[Network Sent (MB)]{
    \begin{tikzpicture}[font=\footnotesize]
        \tkzKiviatDiagram[
            gap=.6,
            scale=.55,
            rotate=90,
            lattice=5,
            label space=.5,
            label style/.style={text width=1em,align=center},
            radial style/.style={draw opacity=0},
        ]{
            $O_1$,
            $O_2$,
            $O_3$,
            $P_{11}$,
            $P_{12}$,
            $P_{21}$,
            $P_{22}$
        }
        \tkzKiviatLineFromFile[linestyle1]{Plots/2mbx2s-network-out.csv}{1}
        \tkzKiviatLineFromFile[linestyle2]{Plots/2mbx2s-network-out.csv}{2}
        \tkzKiviatLineFromFile[linestyle3]{Plots/2mbx2s-network-out.csv}{4}
        \tkzKiviatLineFromFile[linestyle4]{Plots/2mbx2s-network-out.csv}{5}
        \tkzKiviatLineFromFile[linestyle5]{Plots/2mbx2s-network-out.csv}{9}
        \tkzKiviatLineFromFile[linestyle6]{Plots/2mbx2s-network-out.csv}{10}
        \tkzKiviatGrad[graduation distance=-15,unity=140](1)
    \end{tikzpicture}
}\\
\subfloat[Disk Written (MB)]{%
    \begin{tikzpicture}[font=\footnotesize]
        \tkzKiviatDiagram[
            gap=.6,
            scale=.55,
            rotate=90,
            lattice=5,
            label space=.5,
            label style/.style={text width=1em,align=center},
            radial style/.style={draw opacity=0},
        ]{
            $O_1$,
            $O_2$,
            $O_3$,
            $P_{11}$,
            $P_{12}$,
            $P_{21}$,
            $P_{22}$
        }
        \tkzKiviatLineFromFile[linestyle1]{Plots/2mbx2s-disk-write.csv}{1}
        \tkzKiviatLineFromFile[linestyle2]{Plots/2mbx2s-disk-write.csv}{2}
        \tkzKiviatLineFromFile[linestyle3]{Plots/2mbx2s-disk-write.csv}{4}
        \tkzKiviatLineFromFile[linestyle4]{Plots/2mbx2s-disk-write.csv}{5}
        \tkzKiviatLineFromFile[linestyle5]{Plots/2mbx2s-disk-write.csv}{9}
        \tkzKiviatLineFromFile[linestyle6]{Plots/2mbx2s-disk-write.csv}{10}
        \tkzKiviatGrad[graduation distance=-13,unity=1](1)
    \end{tikzpicture}
}
\subfloat[Disk Read (MB)]{
    \begin{tikzpicture}[font=\footnotesize]
        \tkzKiviatDiagram[
            gap=.6,
            scale=.55,
            rotate=90,
            lattice=5,
            label space=.5,
            label style/.style={text width=1em,align=center},
            radial style/.style={draw opacity=0},
        ]{
            $O_1$,
            $O_2$,
            $O_3$,
            $P_{11}$,
            $P_{12}$,
            $P_{21}$,
            $P_{22}$
        }
        \tkzKiviatLineFromFile[linestyle1]{Plots/2mbx2s-disk-read.csv}{1}
        \tkzKiviatLineFromFile[linestyle2]{Plots/2mbx2s-disk-read.csv}{2}
        \tkzKiviatLineFromFile[linestyle3]{Plots/2mbx2s-disk-read.csv}{4}
        \tkzKiviatLineFromFile[linestyle4]{Plots/2mbx2s-disk-read.csv}{5}
        \tkzKiviatLineFromFile[linestyle5]{Plots/2mbx2s-disk-read.csv}{9}
        \tkzKiviatLineFromFile[linestyle6]{Plots/2mbx2s-disk-read.csv}{10}
        \tkzKiviatGrad[graduation distance=-13,unity=1](1)
    \end{tikzpicture}
}

\vspace{0.25cm}
\begin{tikzpicture}
    \matrix [fill=white,draw=black,cells={anchor=west}]{
        \LegendImage{linestyle1} & \LegendEntry{Query a device}; &
        \LegendImage{linestyle2} & \LegendEntry{Query all devices}; &
        \LegendImage{linestyle3} & \LegendEntry{Query a service}; \\
        \LegendImage{linestyle4} & \LegendEntry{Query all services}; &
        \LegendImage{linestyle5} & \LegendEntry{Query a request}; &
        \LegendImage{linestyle6} & \LegendEntry{Query all requests}; \\
    };
\end{tikzpicture}
\caption{System resource utilization of orderer1 ($O_1$), orderer2 ($O_2$), orderer3 ($O_3$), org1 peer1 ($P_{11}$), org1 peer2 ($P_{12}$), org2 peer1 ($P_{21}$), and org2 peer2 ($P_{22}$) during read operations when $S_{soft}=2\rm{MB}$.\label{fig:2mbx2s-read-resource-utilization}}
\end{figure}

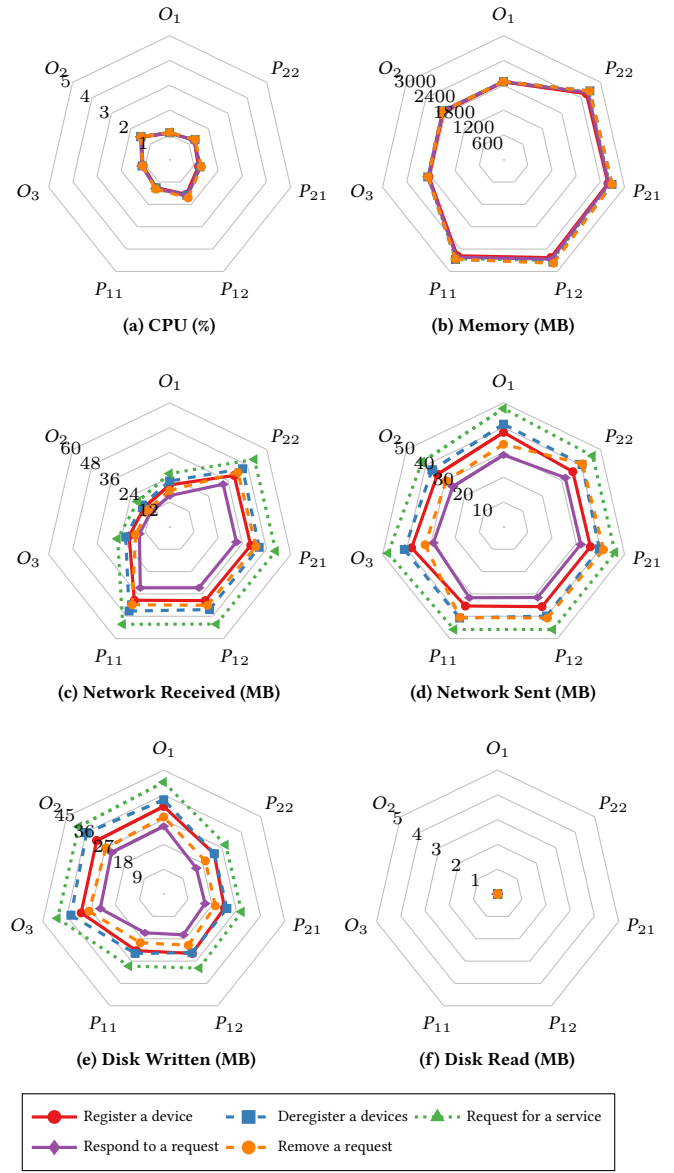
\begin{figure}[htb]
\centering
\captionsetup[subfigure]{justification=centering}
\subfloat[CPU (\%)]{
    \begin{tikzpicture}[font=\footnotesize]
        \tkzKiviatDiagram[
            gap=0.6,
            scale=.55,
            rotate=90,
            lattice=5,
            label space=.5,
            label style/.style={text width=1em,align=center},
            radial style/.style={draw opacity=0},
        ]{
            $O_1$,
            $O_2$,
            $O_3$,
            $P_{11}$,
            $P_{12}$,
            $P_{21}$,
            $P_{22}$
        }
        \tkzKiviatLineFromFile[linestyle1]{Plots/2mbx2s-cpu.csv}{3}
        \tkzKiviatLineFromFile[linestyle2]{Plots/2mbx2s-cpu.csv}{6}
        \tkzKiviatLineFromFile[linestyle3]{Plots/2mbx2s-cpu.csv}{7}
        \tkzKiviatLineFromFile[linestyle4]{Plots/2mbx2s-cpu.csv}{8}
        \tkzKiviatLineFromFile[linestyle5]{Plots/2mbx2s-cpu.csv}{11}
        \tkzKiviatGrad[graduation distance=-15,unity=1](1)
    \end{tikzpicture}
}
\subfloat[Memory (MB)]{
    \begin{tikzpicture}[font=\footnotesize]
        \tkzKiviatDiagram[
            gap=0.6,
            scale=.55,
            rotate=90,
            lattice=5,
            label space=.5,
            label style/.style={text width=1em,align=center},
            radial style/.style={draw opacity=0},
        ]{
            $O_1$,
            $O_2$,
            $O_3$,
            $P_{11}$,
            $P_{12}$,
            $P_{21}$,
            $P_{22}$
        }
        \tkzKiviatLineFromFile[linestyle1]{Plots/2mbx2s-memory.csv}{3}
        \tkzKiviatLineFromFile[linestyle2]{Plots/2mbx2s-memory.csv}{6}
        \tkzKiviatLineFromFile[linestyle3]{Plots/2mbx2s-memory.csv}{7}
        \tkzKiviatLineFromFile[linestyle4]{Plots/2mbx2s-memory.csv}{8}
        \tkzKiviatLineFromFile[linestyle5]{Plots/2mbx2s-memory.csv}{11}
        \tkzKiviatGrad[graduation distance=-15,unity=600](1)
    \end{tikzpicture}
}\\
\subfloat[Network Received (MB)]{
    \begin{tikzpicture}[font=\footnotesize]
        \tkzKiviatDiagram[
            gap=0.6,
            scale=.55,
            rotate=90,
            lattice=5,
            label space=.5,
            label style/.style={text width=1em,align=center},
            radial style/.style={draw opacity=0},
        ]{
            $O_1$,
            $O_2$,
            $O_3$,
            $P_{11}$,
            $P_{12}$,
            $P_{21}$,
            $P_{22}$
        }
        \tkzKiviatLineFromFile[linestyle1]{Plots/2mbx2s-network-in.csv}{3}
        \tkzKiviatLineFromFile[linestyle2]{Plots/2mbx2s-network-in.csv}{6}
        \tkzKiviatLineFromFile[linestyle3]{Plots/2mbx2s-network-in.csv}{7}
        \tkzKiviatLineFromFile[linestyle4]{Plots/2mbx2s-network-in.csv}{8}
        \tkzKiviatLineFromFile[linestyle5]{Plots/2mbx2s-network-in.csv}{11}
        \tkzKiviatGrad[graduation distance=-15,unity=12](1)
    \end{tikzpicture}
}
\subfloat[Network Sent (MB)]{
    \begin{tikzpicture}[font=\footnotesize]
        \tkzKiviatDiagram[
            gap=0.6,
            scale=.55,
            rotate=90,
            lattice=5,
            label space=.5,
            label style/.style={text width=1em,align=center},
            radial style/.style={draw opacity=0},
        ]{
            $O_1$,
            $O_2$,
            $O_3$,
            $P_{11}$,
            $P_{12}$,
            $P_{21}$,
            $P_{22}$
        }
        \tkzKiviatLineFromFile[linestyle1]{Plots/2mbx2s-network-out.csv}{3}
        \tkzKiviatLineFromFile[linestyle2]{Plots/2mbx2s-network-out.csv}{6}
        \tkzKiviatLineFromFile[linestyle3]{Plots/2mbx2s-network-out.csv}{7}
        \tkzKiviatLineFromFile[linestyle4]{Plots/2mbx2s-network-out.csv}{8}
        \tkzKiviatLineFromFile[linestyle5]{Plots/2mbx2s-network-out.csv}{11}
        \tkzKiviatGrad[graduation distance=-15,unity=10](1)
    \end{tikzpicture}
}\\
\subfloat[Disk Written (MB)]{%
    \begin{tikzpicture}[font=\footnotesize]
        \tkzKiviatDiagram[
            gap=0.6,
            scale=.55,
            rotate=90,
            lattice=5,
            label space=.5,
            label style/.style={text width=1em,align=center},
            radial style/.style={draw opacity=0},
        ]{
            $O_1$,
            $O_2$,
            $O_3$,
            $P_{11}$,
            $P_{12}$,
            $P_{21}$,
            $P_{22}$
        }
        \tkzKiviatLineFromFile[linestyle1]{Plots/2mbx2s-disk-write.csv}{3}
        \tkzKiviatLineFromFile[linestyle2]{Plots/2mbx2s-disk-write.csv}{6}
        \tkzKiviatLineFromFile[linestyle3]{Plots/2mbx2s-disk-write.csv}{7}
        \tkzKiviatLineFromFile[linestyle4]{Plots/2mbx2s-disk-write.csv}{8}
        \tkzKiviatLineFromFile[linestyle5]{Plots/2mbx2s-disk-write.csv}{11}
        \tkzKiviatGrad[graduation distance=-15,unity=9](1)
    \end{tikzpicture}
}
\subfloat[Disk Read (MB)]{
    \begin{tikzpicture}[font=\footnotesize]
        \tkzKiviatDiagram[
            gap=0.6,
            scale=.55,
            rotate=90,
            lattice=5,
            label space=.5,
            label style/.style={text width=1em,align=center},
            radial style/.style={draw opacity=0},
        ]{
            $O_1$,
            $O_2$,
            $O_3$,
            $P_{11}$,
            $P_{12}$,
            $P_{21}$,
            $P_{22}$
        }
        \tkzKiviatLineFromFile[linestyle1]{Plots/2mbx2s-disk-read.csv}{3}
        \tkzKiviatLineFromFile[linestyle2]{Plots/2mbx2s-disk-read.csv}{6}
        \tkzKiviatLineFromFile[linestyle3]{Plots/2mbx2s-disk-read.csv}{7}
        \tkzKiviatLineFromFile[linestyle4]{Plots/2mbx2s-disk-read.csv}{8}
        \tkzKiviatLineFromFile[linestyle5]{Plots/2mbx2s-disk-read.csv}{11}
        \tkzKiviatGrad[graduation distance=-13,unity=1](1)
    \end{tikzpicture}
}

\vspace{0.25cm}
\begin{tikzpicture}
    \matrix [fill=white,draw=black,cells={anchor=west}]{
        \LegendImage{linestyle1} & \LegendEntry{Register a device}; &
        \LegendImage{linestyle2} & \LegendEntry{Deregister a devices}; &
        \LegendImage{linestyle3} & \LegendEntry{Request for a service}; \\
        \LegendImage{linestyle4} & \LegendEntry{Respond to a request}; &
        \LegendImage{linestyle5} & \LegendEntry{Remove a request}; \\
    };
\end{tikzpicture}
\caption{System resource utilization of orderer1 ($O_1$), orderer2 ($O_2$), orderer3 ($O_3$), org1 peer1 ($P_{11}$), org1 peer2 ($P_{12}$), org2 peer1 ($P_{21}$), and org2 peer2 ($P_{22}$) during transaction operations when $S_{soft}=2\rm{MB}$.\label{fig:2mbx2s-transaction-resource-utilization}}
\end{figure}

Before investigating batch size's impact on system resource utilization of smart contract operations, we first explore the similarities in resource utilization patterns of different operations.
Figure~\ref{fig:2mbx2s-read-resource-utilization} and Figure~\ref{fig:2mbx2s-transaction-resource-utilization} present the average CPU, average memory, total network, and total disk usage of every host of the test Hyperledger Fabric network with 2MB batches.
The results show that peer nodes are more utilized than orderer nodes under most workloads.
And the utilization levels are even on the same type of nodes.
Regardless of the specific operation performed, all read operations show similar resource usage patterns, and so are transaction operations.
Meanwhile, transaction operations involve more incoming network traffic and disk writes than read operations, and operations that consume and produce larger messages require more CPU and memory.
Finally, disks are rarely read in all tests due to ledger data being loaded to memory before the measurements begin.

The same resource utilization patterns are also observed in other tests with varying batch sizes.
Therefore, we only present the results of the ``request for a service'' and ``query a request'' operations to demonstrate the relationship between resource utilization and batch sizes.
Figure~\ref{fig:utilization-read-vs-batch-size} and Figure~\ref{fig:utilization-transaction-vs-batch-size} depict the CPU, memory, network, and disk usage of each test given each batch size.
For read operations such as querying a single service request, there is slight fluctuation in system resource usage as the batch size increases, just like their throughput and latency.
This result aligns with the fact that those operations do not go through the ordering process.
On the other hand, the transaction operations incur higher CPU utilization, and more disk writes when the batch size is small.
It is clear that with smaller batch sizes, the network needs to create more blocks to commit the same number of transactions, hence, more overhead in the data to be written to disk and the computing power required to complete that.
As for network usage, it does not change significantly with batch sizes because the size of incoming and outgoing messages is the same regardless of batch sizes.

\begin{figure}[htb]
\centering
\pgfplotsset{
    xmin=0,
    ymin=0,
    width=.9\linewidth,
    height=1.75in,
    xlabel={Batch Size (MB)},
    xtick distance=1,
    legend columns=4,
    legend cell align={left},
    legend style={at={(0.5,-0.3)},anchor=north},
}
\captionsetup[subfigure]{justification=centering}
\subfloat[CPU and Memory Utilization]{
    \begin{tikzpicture}
    \begin{axis}[
        ylabel={CPU (\%)},
        ytick distance=0.2,
        axis y line*=left,
    ]
        \addplot+ table [x={Batch Size (MB)}, y={Avg CPU Orderer}, col sep=tab] {Plots/utilization-read-vs-batch-size.csv};
        \label{orderer-cpu-read-batch-size}
        \addplot+ table [x={Batch Size (MB)}, y={Avg CPU Peer}, col sep=tab] {Plots/utilization-read-vs-batch-size.csv};
        \label{peer-cpu-read-batch-size}
    \end{axis}
    \begin{axis}[
        axis x line=none,
        axis y line*=right,
        ylabel={Memory (MB)},
        ytick distance=400,
    ]
        \addlegendimage{/pgfplots/refstyle=orderer-cpu-read-batch-size}
        \addlegendentry{Orderer CPU}
        \addlegendimage{/pgfplots/refstyle=peer-cpu-read-batch-size}
        \addlegendentry{Peer CPU}
        \addplot+[linestyle3] table [x={Batch Size (MB)}, y={Avg Memory Orderer (MB)}, col sep=tab] {Plots/utilization-read-vs-batch-size.csv};
        \addlegendentry{Orderer memory}
        \addplot+[linestyle4] table [x={Batch Size (MB)}, y={Avg Memory Peer (MB)}, col sep=tab] {Plots/utilization-read-vs-batch-size.csv};
        \addlegendentry{Peer memory}
    \end{axis}
    \end{tikzpicture}
}\\
\subfloat[Network Utilization]{
    \begin{tikzpicture}
    \begin{axis}[
        ylabel={Sent/Received (MB)},
        ytick distance=0.5,
    ]
        \addplot+ table [x={Batch Size (MB)}, y={Avg Network In Orderer (MB)}, col sep=tab] {Plots/utilization-read-vs-batch-size.csv};
        \addlegendentry{Orderer received}
        \addplot+ table [x={Batch Size (MB)}, y={Avg Network In Peer (MB)}, col sep=tab] {Plots/utilization-read-vs-batch-size.csv};
        \addlegendentry{Peer received}
        \addplot+ table [x={Batch Size (MB)}, y={Avg Network Out Orderer (MB)}, col sep=tab] {Plots/utilization-read-vs-batch-size.csv};
        \addlegendentry{Orderer sent}
        \addplot+ table [x={Batch Size (MB)}, y={Avg Network Out Peer (MB)}, col sep=tab] {Plots/utilization-read-vs-batch-size.csv};
        \addlegendentry{Peer sent}
    \end{axis}
    \end{tikzpicture}
}\\
\subfloat[Disk Utilization]{
    \begin{tikzpicture}
    \begin{axis}[
        ymax=2,
        ylabel={Read/Write (MB)},
        ytick distance=0.4,
    ]
        \addplot+ table [x={Batch Size (MB)}, y={Avg Disk Write Orderer (MB)}, col sep=tab] {Plots/utilization-read-vs-batch-size.csv};
        \addlegendentry{Orderer write}
        \addplot+ table [x={Batch Size (MB)}, y={Avg Disk Write Peer (MB)}, col sep=tab] {Plots/utilization-read-vs-batch-size.csv};
        \addlegendentry{Peer write}
        \addplot+ table [x={Batch Size (MB)}, y={Avg Disk Read Orderer (MB)}, col sep=tab] {Plots/utilization-read-vs-batch-size.csv};
        \addlegendentry{Orderer read}
        \addplot+ table [x={Batch Size (MB)}, y={Avg Disk Read Peer (MB)}, col sep=tab] {Plots/utilization-read-vs-batch-size.csv};
        \addlegendentry{Peer read}
    \end{axis}
    \end{tikzpicture}
}
\caption{Average system resource utilization of orderer and peer nodes for handling ``querying a service request'' operations under varying batch sizes.\label{fig:utilization-read-vs-batch-size}}
\end{figure}

\begin{figure}[htb]
\centering
\pgfplotsset{
    xmin=0,
    ymin=0,
    width=.9\linewidth,
    height=1.75in,
    xlabel={Batch Size (MB)},
    xtick distance=1,
    legend columns=4,
    legend cell align={left},
    legend style={at={(0.5,-0.3)},anchor=north},
}
\captionsetup[subfigure]{justification=centering}
\subfloat[CPU and Memory Utilization]{
    \begin{tikzpicture}
    \begin{axis}[
        ylabel={CPU (\%)},
        ytick distance=0.4,
        axis y line*=left,
    ]
        \addplot+ table [x={Batch Size (MB)}, y={Avg CPU Orderer}, col sep=tab] {Plots/utilization-transaction-vs-batch-size.csv};
        \label{orderer-cpu-transaction-batch-size}
        \addplot+ table [x={Batch Size (MB)}, y={Avg CPU Peer}, col sep=tab] {Plots/utilization-transaction-vs-batch-size.csv};
        \label{peer-cpu-transaction-batch-size}
    \end{axis}
    \begin{axis}[
        axis x line=none,
        axis y line*=right,
        ylabel={Memory (MB)},
        ytick distance=400,
    ]
        \addlegendimage{/pgfplots/refstyle=orderer-cpu-transaction-batch-size}
        \addlegendentry{Orderer CPU}
        \addlegendimage{/pgfplots/refstyle=peer-cpu-transaction-batch-size}
        \addlegendentry{Peer CPU}
        \addplot+[linestyle3] table [x={Batch Size (MB)}, y={Avg Memory Orderer (MB)}, col sep=tab] {Plots/utilization-transaction-vs-batch-size.csv};
        \addlegendentry{Orderer memory}
        \addplot+[linestyle4] table [x={Batch Size (MB)}, y={Avg Memory Peer (MB)}, col sep=tab] {Plots/utilization-transaction-vs-batch-size.csv};
        \addlegendentry{Peer memory}
    \end{axis}
    \end{tikzpicture}
}\\
\subfloat[Network Utilization]{
    \begin{tikzpicture}
    \begin{axis}[
        ylabel={Sent/Received (MB)},
        ytick distance=10,
    ]
        \addplot+ table [x={Batch Size (MB)}, y={Avg Network In Orderer (MB)}, col sep=tab] {Plots/utilization-transaction-vs-batch-size.csv};
        \addlegendentry{Orderer received}
        \addplot+ table [x={Batch Size (MB)}, y={Avg Network In Peer (MB)}, col sep=tab] {Plots/utilization-transaction-vs-batch-size.csv};
        \addlegendentry{Peer received}
        \addplot+ table [x={Batch Size (MB)}, y={Avg Network Out Orderer (MB)}, col sep=tab] {Plots/utilization-transaction-vs-batch-size.csv};
        \addlegendentry{Orderer sent}
        \addplot+ table [x={Batch Size (MB)}, y={Avg Network Out Peer (MB)}, col sep=tab] {Plots/utilization-transaction-vs-batch-size.csv};
        \addlegendentry{Peer sent}
    \end{axis}
    \end{tikzpicture}
}\\
\subfloat[Disk Utilization]{
    \begin{tikzpicture}
    \begin{axis}[
        ylabel={Read/Write (MB)},
        ytick distance=10,
    ]
        \addplot+ table [x={Batch Size (MB)}, y={Avg Disk Write Orderer (MB)}, col sep=tab] {Plots/utilization-transaction-vs-batch-size.csv};
        \addlegendentry{Orderer write}
        \addplot+ table [x={Batch Size (MB)}, y={Avg Disk Write Peer (MB)}, col sep=tab] {Plots/utilization-transaction-vs-batch-size.csv};
        \addlegendentry{Peer write}
        \addplot+ table [x={Batch Size (MB)}, y={Avg Disk Read Orderer (MB)}, col sep=tab] {Plots/utilization-transaction-vs-batch-size.csv};
        \addlegendentry{Orderer read}
        \addplot+ table [x={Batch Size (MB)}, y={Avg Disk Read Peer (MB)}, col sep=tab] {Plots/utilization-transaction-vs-batch-size.csv};
        \addlegendentry{Peer read}
    \end{axis}
    \end{tikzpicture}
}
\caption{Average system resource utilization of orderer and peer nodes for handling ``requesting for service''  operations under varying batch sizes.\label{fig:utilization-transaction-vs-batch-size}}
\end{figure}

\subsection{Performance vs. Batch Timeout}

\begin{figure}[htb]
\centering
\begin{tikzpicture}
    \begin{groupplot}[
        group style={group size=1 by 2},
        xmin=0,
        ymin=0,
        width=.9\linewidth,
        height=1.75in,
        xlabel={Batch Timeout (s)},
        xtick distance=1,
    ]
    \nextgroupplot[ylabel={Throughput (TPS)}]
        \addplot+ table [x={Batch Timeout (s)}, y={/service-registry/register}, col sep=tab] {Plots/throughput-vs-batch-timeout.csv};
        \addplot+ table [x={Batch Timeout (s)}, y={/service-registry/deregister}, col sep=tab] {Plots/throughput-vs-batch-timeout.csv};
        \addplot+ table [x={Batch Timeout (s)}, y={/service-broker/request}, col sep=tab] {Plots/throughput-vs-batch-timeout.csv};
        \addplot+ table [x={Batch Timeout (s)}, y={/service-broker/respond}, col sep=tab] {Plots/throughput-vs-batch-timeout.csv};
        \addplot+ table [x={Batch Timeout (s)}, y={/service-broker/remove}, col sep=tab] {Plots/throughput-vs-batch-timeout.csv};
    
    \nextgroupplot[ylabel={Latency (s)},ytick distance=1]
        \addplot+ table [x={Batch Timeout (s)}, y={/service-registry/register}, col sep=tab] {Plots/latency-vs-batch-timeout.csv};
        \addplot+ table [x={Batch Timeout (s)}, y={/service-registry/deregister}, col sep=tab] {Plots/latency-vs-batch-timeout.csv};
        \addplot+ table [x={Batch Timeout (s)}, y={/service-broker/request}, col sep=tab] {Plots/latency-vs-batch-timeout.csv};
        \addplot+ table [x={Batch Timeout (s)}, y={/service-broker/respond}, col sep=tab] {Plots/latency-vs-batch-timeout.csv};
        \addplot+ table [x={Batch Timeout (s)}, y={/service-broker/remove}, col sep=tab] {Plots/latency-vs-batch-timeout.csv};
    \end{groupplot}
\end{tikzpicture}

\vspace{0.25cm}
\begin{tikzpicture}
    \matrix [fill=white,draw=black,cells={anchor=west}]{
        \LegendImage{linestyle1} & \LegendEntry{Register a device}; &
        \LegendImage{linestyle2} & \LegendEntry{Deregister a devices}; &
        \LegendImage{linestyle3} & \LegendEntry{Request for a service}; \\
        \LegendImage{linestyle4} & \LegendEntry{Respond to a request}; &
        \LegendImage{linestyle5} & \LegendEntry{Remove a request}; \\
    };
\end{tikzpicture}
\caption{Throughput and latency of transaction operations for varying batch timeout.\label{fig:performance-transaction-vs-batch-timeout}}
\end{figure}
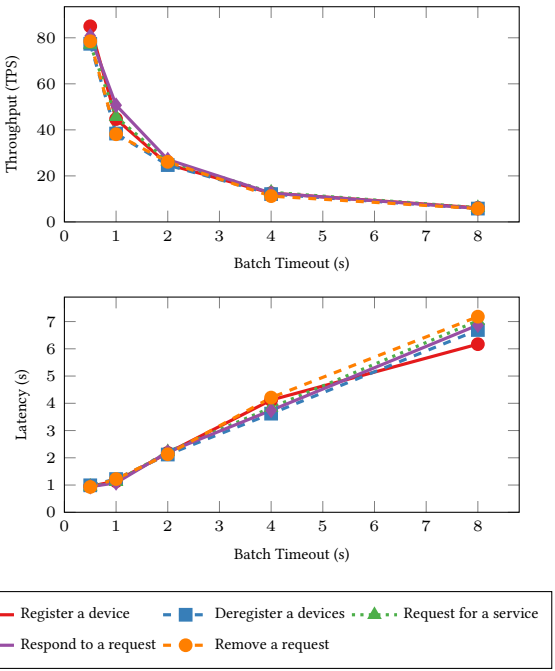

Batch timeout $T_{timeout}$ is another essential configuration that controls the generation of blocks.
It is the maximum time to wait before creating a new block after the first transaction arrives at the ordering service.
To estimate the impact of batch timeout on the performance of the proposed system, we perform the same experiments as the previous section, with batch timeouts set to 500ms, 1s, 2s, 4s, and 8s.
The throughput and latency of the read and transaction operations are shown in Figure~\ref{fig:performance-read-vs-batch-timeout} and Figure~\ref{fig:performance-transaction-vs-batch-timeout}.
Similar to the results of batch size tests, the throughput and latency of read operations are unaffected by the change of batch timeout because the ordering service does not process such operations.
Transaction operations show a logarithmic decrease in throughput and a linear increase in latency regarding batch timeout.
Blocks wait longer to be created when timeout is large, and the commit of transactions is delayed as a result.
However, setting a large timeout also allows more transactions to be batched into the same block, mitigating the decrease in throughput.

\begin{figure}[htb]
\centering
\begin{tikzpicture}
    \begin{groupplot}[
        group style={group size=1 by 2},
        xmin=0,
        ymin=0,
        width=.9\linewidth,
        height=1.75in,
        xlabel={Batch Timeout (s)},
        xtick distance=1,
    ]
    \nextgroupplot[ylabel={Throughput (RPS)}]
        \addplot+ table [x={Batch Timeout (s)}, y={/device-registry/get}, col sep=tab] {Plots/throughput-vs-batch-timeout.csv};
        \addplot+ table [x={Batch Timeout (s)}, y={/device-registry/get-all}, col sep=tab] {Plots/throughput-vs-batch-timeout.csv};
        \addplot+ table [x={Batch Timeout (s)}, y={/service-registry/get}, col sep=tab] {Plots/throughput-vs-batch-timeout.csv};
        \addplot+ table [x={Batch Timeout (s)}, y={/service-registry/get-all}, col sep=tab] {Plots/throughput-vs-batch-timeout.csv};
        \addplot+ table [x={Batch Timeout (s)}, y={/service-broker/get}, col sep=tab] {Plots/throughput-vs-batch-timeout.csv};
        \addplot+ table [x={Batch Timeout (s)}, y={/service-broker/get-all}, col sep=tab] {Plots/throughput-vs-batch-timeout.csv};
    
    \nextgroupplot[ylabel={Latency (s)}, ytick distance=3]
        \addplot+ table [x={Batch Timeout (s)}, y={/device-registry/get}, col sep=tab] {Plots/latency-vs-batch-timeout.csv};
        \addplot+ table [x={Batch Timeout (s)}, y={/device-registry/get-all}, col sep=tab] {Plots/latency-vs-batch-timeout.csv};
        \addplot+ table [x={Batch Timeout (s)}, y={/service-registry/get}, col sep=tab] {Plots/latency-vs-batch-timeout.csv};
        \addplot+ table [x={Batch Timeout (s)}, y={/service-registry/get-all}, col sep=tab] {Plots/latency-vs-batch-timeout.csv};
        \addplot+ table [x={Batch Timeout (s)}, y={/service-broker/get}, col sep=tab] {Plots/latency-vs-batch-timeout.csv};
        \addplot+ table [x={Batch Timeout (s)}, y={/service-broker/get-all}, col sep=tab] {Plots/latency-vs-batch-timeout.csv};
    \end{groupplot}
\end{tikzpicture}

\vspace{0.25cm}
\begin{tikzpicture}
    \matrix [fill=white,draw=black,cells={anchor=west}]{
        \LegendImage{linestyle1} & \LegendEntry{Query a device}; &
        \LegendImage{linestyle2} & \LegendEntry{Query all devices}; &
        \LegendImage{linestyle3} & \LegendEntry{Query a service}; \\
        \LegendImage{linestyle4} & \LegendEntry{Query all services}; &
        \LegendImage{linestyle5} & \LegendEntry{Query a request}; &
        \LegendImage{linestyle6} & \LegendEntry{Query all requests}; \\
    };
\end{tikzpicture}
\caption{Throughput and latency of read operations for varying batch timeout.\label{fig:performance-read-vs-batch-timeout}}
\end{figure}
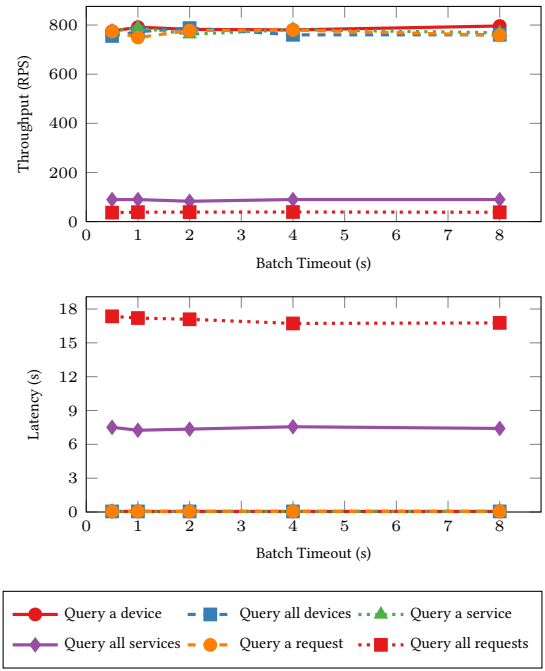

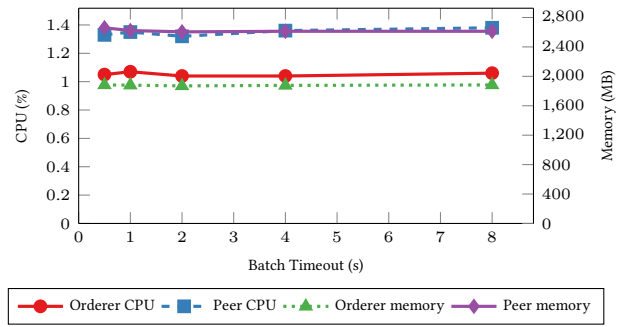
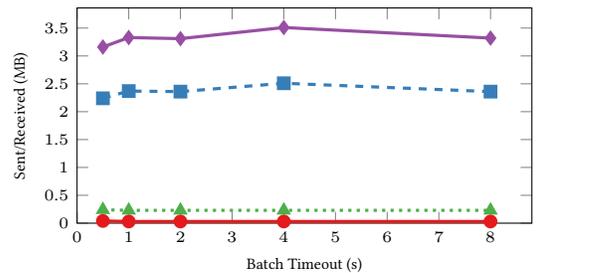
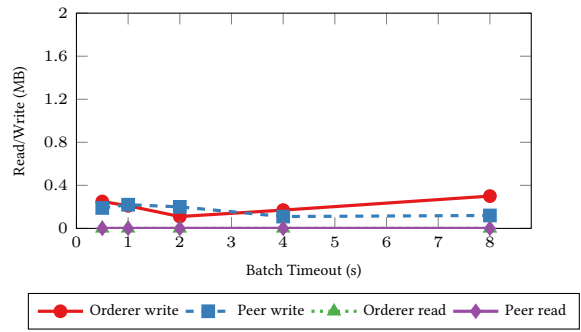
\begin{figure}[t]
\centering
\pgfplotsset{
    xmin=0,
    ymin=0,
    width=.9\linewidth,
    height=1.75in,
    xlabel={Batch Timeout (s)},
    xtick distance=1,
    legend columns=4,
    legend cell align={left},
    legend style={at={(0.5,-0.3)},anchor=north},
}
\captionsetup[subfigure]{justification=centering}
\subfloat[CPU and Memory Utilization]{
    \begin{tikzpicture}
    \begin{axis}[
        ylabel={CPU (\%)},
        ytick distance=0.2,
        axis y line*=left,
    ]
        \addplot+ table [x={Batch Timeout (s)}, y={Avg CPU Orderer}, col sep=tab] {Plots/utilization-read-vs-batch-timeout.csv};
        \label{orderer-cpu-read-batch-timeout}
        \addplot+ table [x={Batch Timeout (s)}, y={Avg CPU Peer}, col sep=tab] {Plots/utilization-read-vs-batch-timeout.csv};
        \label{peer-cpu-read-batch-timeout}
    \end{axis}
    \begin{axis}[
        axis x line=none,
        axis y line*=right,
        ylabel={Memory (MB)},
        ytick distance=400,
    ]
        \addlegendimage{/pgfplots/refstyle=orderer-cpu-read-batch-timeout}
        \addlegendentry{Orderer CPU}
        \addlegendimage{/pgfplots/refstyle=peer-cpu-read-batch-timeout}
        \addlegendentry{Peer CPU}
        \addplot+[linestyle3] table [x={Batch Timeout (s)}, y={Avg Memory Orderer (MB)}, col sep=tab] {Plots/utilization-read-vs-batch-timeout.csv};
        \addlegendentry{Orderer memory}
        \addplot+[linestyle4] table [x={Batch Timeout (s)}, y={Avg Memory Peer (MB)}, col sep=tab] {Plots/utilization-read-vs-batch-timeout.csv};
        \addlegendentry{Peer memory}
    \end{axis}
    \end{tikzpicture}
}\\
\subfloat[Network Utilization]{
    \begin{tikzpicture}
    \begin{axis}[
        ylabel={Sent/Received (MB)},
        ytick distance=0.5,
    ]
        \addplot+ table [x={Batch Timeout (s)}, y={Avg Network In Orderer (MB)}, col sep=tab] {Plots/utilization-read-vs-batch-timeout.csv};
        \addlegendentry{Orderer received}
        \addplot+ table [x={Batch Timeout (s)}, y={Avg Network In Peer (MB)}, col sep=tab] {Plots/utilization-read-vs-batch-timeout.csv};
        \addlegendentry{Peer received}
        \addplot+ table [x={Batch Timeout (s)}, y={Avg Network Out Orderer (MB)}, col sep=tab] {Plots/utilization-read-vs-batch-timeout.csv};
        \addlegendentry{Orderer sent}
        \addplot+ table [x={Batch Timeout (s)}, y={Avg Network Out Peer (MB)}, col sep=tab] {Plots/utilization-read-vs-batch-timeout.csv};
        \addlegendentry{Peer sent}
    \end{axis}
    \end{tikzpicture}
}\\
\subfloat[Disk Utilization]{
    \begin{tikzpicture}
    \begin{axis}[
        ymax=2,
        ylabel={Read/Write (MB)},
        ytick distance=0.4,
    ]
        \addplot+ table [x={Batch Timeout (s)}, y={Avg Disk Write Orderer (MB)}, col sep=tab] {Plots/utilization-read-vs-batch-timeout.csv};
        \addlegendentry{Orderer write}
        \addplot+ table [x={Batch Timeout (s)}, y={Avg Disk Write Peer (MB)}, col sep=tab] {Plots/utilization-read-vs-batch-timeout.csv};
        \addlegendentry{Peer write}
        \addplot+ table [x={Batch Timeout (s)}, y={Avg Disk Read Orderer (MB)}, col sep=tab] {Plots/utilization-read-vs-batch-timeout.csv};
        \addlegendentry{Orderer read}
        \addplot+ table [x={Batch Timeout (s)}, y={Avg Disk Read Peer (MB)}, col sep=tab] {Plots/utilization-read-vs-batch-timeout.csv};
        \addlegendentry{Peer read}
    \end{axis}
    \end{tikzpicture}
}
\caption{Average system resource utilization of orderer and peer nodes for handling ``querying a service request'' operations under varying batch timeouts.\label{fig:utilization-read-vs-batch-timeout}}
\end{figure}

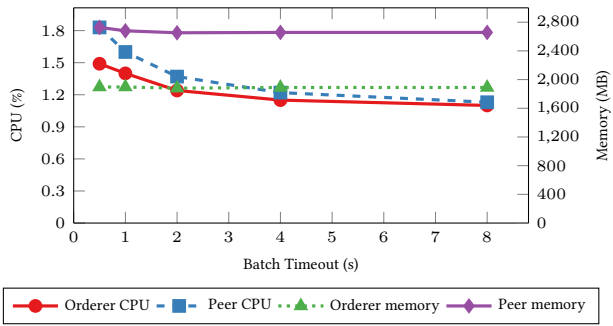
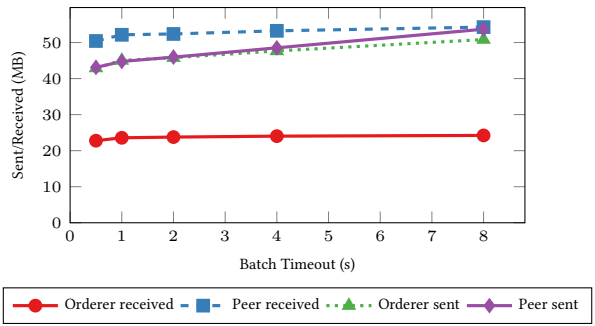
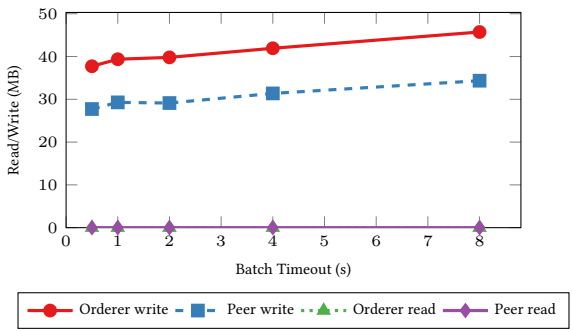
\begin{figure}[!t]
\centering
\pgfplotsset{
    xmin=0,
    ymin=0,
    width=.9\linewidth,
    height=1.75in,
    xlabel={Batch Timeout (s)},
    xtick distance=1,
    legend columns=4,
    legend cell align={left},
    legend style={at={(0.5,-0.3)},anchor=north},
}
\captionsetup[subfigure]{justification=centering}
\subfloat[CPU and Memory Utilization]{
    \begin{tikzpicture}
    \begin{axis}[
        ylabel={CPU (\%)},
        ytick distance=0.3,
        axis y line*=left,
    ]
        \addplot+ table [x={Batch Timeout (s)}, y={Avg CPU Orderer}, col sep=tab] {Plots/utilization-transaction-vs-batch-timeout.csv};
        \label{orderer-cpu-transaction-batch-timeout}
        \addplot+ table [x={Batch Timeout (s)}, y={Avg CPU Peer}, col sep=tab] {Plots/utilization-transaction-vs-batch-timeout.csv};
        \label{peer-cpu-transaction-batch-timeout}
    \end{axis}
    \begin{axis}[
        axis x line=none,
        axis y line*=right,
        ylabel={Memory (MB)},
        ytick distance=400,
    ]
        \addlegendimage{/pgfplots/refstyle=orderer-cpu-transaction-batch-timeout}
        \addlegendentry{Orderer CPU}
        \addlegendimage{/pgfplots/refstyle=peer-cpu-transaction-batch-timeout}
        \addlegendentry{Peer CPU}
        \addplot+[linestyle3] table [x={Batch Timeout (s)}, y={Avg Memory Orderer (MB)}, col sep=tab] {Plots/utilization-transaction-vs-batch-timeout.csv};
        \addlegendentry{Orderer memory}
        \addplot+[linestyle4] table [x={Batch Timeout (s)}, y={Avg Memory Peer (MB)}, col sep=tab] {Plots/utilization-transaction-vs-batch-timeout.csv};
        \addlegendentry{Peer memory}
    \end{axis}
    \end{tikzpicture}
}\\
\subfloat[Network Utilization]{
    \begin{tikzpicture}
    \begin{axis}[
        ylabel={Sent/Received (MB)},
        ytick distance=10,
    ]
        \addplot+ table [x={Batch Timeout (s)}, y={Avg Network In Orderer (MB)}, col sep=tab] {Plots/utilization-transaction-vs-batch-timeout.csv};
        \addlegendentry{Orderer received}
        \addplot+ table [x={Batch Timeout (s)}, y={Avg Network In Peer (MB)}, col sep=tab] {Plots/utilization-transaction-vs-batch-timeout.csv};
        \addlegendentry{Peer received}
        \addplot+ table [x={Batch Timeout (s)}, y={Avg Network Out Orderer (MB)}, col sep=tab] {Plots/utilization-transaction-vs-batch-timeout.csv};
        \addlegendentry{Orderer sent}
        \addplot+ table [x={Batch Timeout (s)}, y={Avg Network Out Peer (MB)}, col sep=tab] {Plots/utilization-transaction-vs-batch-timeout.csv};
        \addlegendentry{Peer sent}
    \end{axis}
    \end{tikzpicture}
}\\
\subfloat[Disk Utilization]{
    \begin{tikzpicture}
    \begin{axis}[
        ylabel={Read/Write (MB)},
        ytick distance=10,
    ]
        \addplot+ table [x={Batch Timeout (s)}, y={Avg Disk Write Orderer (MB)}, col sep=tab] {Plots/utilization-transaction-vs-batch-timeout.csv};
        \addlegendentry{Orderer write}
        \addplot+ table [x={Batch Timeout (s)}, y={Avg Disk Write Peer (MB)}, col sep=tab] {Plots/utilization-transaction-vs-batch-timeout.csv};
        \addlegendentry{Peer write}
        \addplot+ table [x={Batch Timeout (s)}, y={Avg Disk Read Orderer (MB)}, col sep=tab] {Plots/utilization-transaction-vs-batch-timeout.csv};
        \addlegendentry{Orderer read}
        \addplot+ table [x={Batch Timeout (s)}, y={Avg Disk Read Peer (MB)}, col sep=tab] {Plots/utilization-transaction-vs-batch-timeout.csv};
        \addlegendentry{Peer read}
    \end{axis}
    \end{tikzpicture}
}
\caption{Average system resource utilization of orderer and peer nodes for handling ``requesting for service'' operations under varying batch timeouts.\label{fig:utilization-transaction-vs-batch-timeout}}
\end{figure}

Figure~\ref{fig:utilization-read-vs-batch-timeout} and Figure~\ref{fig:utilization-transaction-vs-batch-timeout} depict the average system resource utilization of orderer and peer nodes for the tested batch timeouts.
Like the previous experiments, read operations exhibit consistent resource usage given different batch timeouts.
Transaction operations, on the other hand, consume more orderer and peer CPU for small batch timeouts.
The CPU usage decreases in the same fashion as the throughput as timeout increases.
We also observe a steady climb in outgoing network traffic and disk write as timeout increases.
Other resources, such as memory and incoming network traffic, are much less affected by the batch timeout.

\subsection{Performance vs. Connection Size}

A real-world IoT network can be composed of hundreds of devices that actively provide services simultaneously.
The increase in device connections inevitably affects the responsiveness of our proposed platform.
Therefore, our final evaluation focuses on the impact of client connection size on our proposed blockchain platform's performance and system resource utilization.
We created 2,000 client connections to the platform using Hyperledger Caliper in order to simulate the situation where a massive number of IoT devices communicate through the platform concurrently.
Then, we rerun the tests with the batch size configured to 2MB and batch time to 2s.
The results are compared with previous results that employ ten client connections to discover the mass connections' performance and resource utilization overhead.

\begin{figure}[htb]
\centering
\begin{tikzpicture}
    \begin{groupplot}[
        group style={
            group size=2 by 2,
            horizontal sep=1em,
            vertical sep=1.2cm,
        },
        width=0.5\linewidth,
        xbar,
        xmin=0,
        ytick=data,
        ytick pos=left,
        y dir=reverse,
        cycle list/Set1-5,
        cycle list name=Set1-5,
        enlarge y limits={abs=0.3cm},
        every axis plot/.append style={fill},
        /pgf/bar shift={
            0.5*(\numplotsofactualtype*\pgfplotbarwidth + (\numplotsofactualtype-1)*(2pt))  - 
            (.5+\plotnumofactualtype)*\pgfplotbarwidth - \plotnumofactualtype*(2pt)
        },
    ]
    \nextgroupplot[
        xlabel={Throughput (RPS)},
        height=3.2cm,
        symbolic y coords={
            /device-registry/get,
            /service-registry/get,
            /service-broker/get,
        },
        yticklabels={
            Query a device,
            Query a service,
            Query a request,
        },
    ]
        \addplot+ table [x index=1, y index=0, col sep=tab] {Plots/throughput-read-vs-connection.csv};
        \addplot+ table [x index=2, y index=0, col sep=tab] {Plots/throughput-read-vs-connection.csv};
    
    \nextgroupplot[
        xlabel={Latency (s)},
        height=3.2cm,
        symbolic y coords={
            /device-registry/get,
            /service-registry/get,
            /service-broker/get,
        },
        yticklabel=\empty,
        scaled ticks=false,
        x tick label style={/pgf/number format/fixed},
    ]
        \addplot+ table [x index=1, y index=0, col sep=tab] {Plots/latency-read-vs-connection.csv};
        \addplot+ table [x index=2, y index=0, col sep=tab] {Plots/latency-read-vs-connection.csv};
        
    \nextgroupplot[
        xlabel={Throughput (TPS)},
        height=4.2cm,
        symbolic y coords={
            /service-registry/register,
            /service-registry/deregister,
            /service-broker/request,
            /service-broker/respond,
            /service-broker/remove,
        },
        yticklabels={
            Register a service,
            Deregister a service,
            Request for a service,
            Respond to a requet,
            Remove a request,
        },
    ]
        \addplot+ table [x index=1, y index=0, col sep=tab] {Plots/throughput-transaction-vs-connection.csv};
        \addplot+ table [x index=2, y index=0, col sep=tab] {Plots/throughput-transaction-vs-connection.csv};
    
    \nextgroupplot[
        xlabel={Latency (s)},
        height=4.2cm,
        symbolic y coords={
            /service-registry/register,
            /service-registry/deregister,
            /service-broker/request,
            /service-broker/respond,
            /service-broker/remove,
        },
        yticklabel=\empty,
        legend columns=2,
        legend cell align={left},
        legend style={at={(-0.45,-0.35)},anchor=north},
        legend image code/.code={\draw [#1] (0cm,-0.075cm) rectangle (0.3cm,0.075cm);},
    ]
        \addplot+ table [x index=1, y index=0, col sep=tab] {Plots/latency-transaction-vs-connection.csv};
        \addlegendentry{$10x$}
        \addplot+ table [x index=2, y index=0, col sep=tab] {Plots/latency-transaction-vs-connection.csv};
        \addlegendentry{$2000x$}
    \end{groupplot}
\end{tikzpicture}
\caption{Throughput and latency of read and transaction operations for 10 client connections ($10x$) and 2,000 client connections ($2,000x$).\label{fig:performance-vs-connection}}
\end{figure}
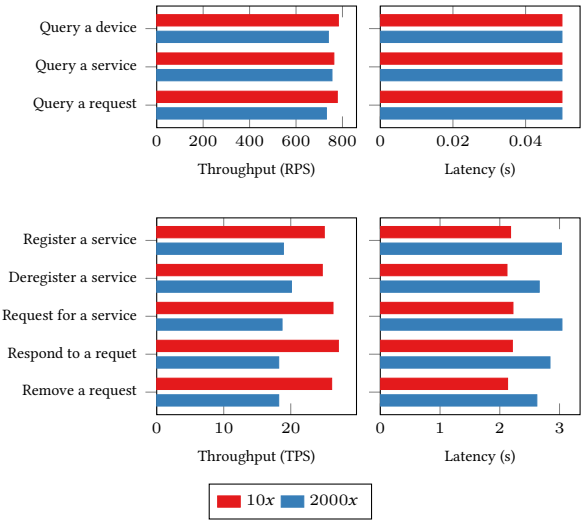

Figure~\ref{fig:performance-vs-connection} presents the throughput and latency of certain operations under two connection size settings.
Batch queries such as ``query all services'' are excluded from the evaluation because the number of devices or services they query is different under the two testing scenarios, so their results are incomparable.
For read operations, the throughput experiences a 5 to 6\% decrease in the 2,000 connections scenario while the latency remains the same.
As for transaction operations, the throughput suffers a 30\% degradation, and latency increases by 25\% to 38\%.
This notable overhead results from massive clients collecting and submitting endorsements from peers.
The more connection established to the peers and orderers, the more congestion there will be hindering transaction processing.
Therefore, it is advised to limit the number of client connections to each peer.

\begin{figure}[p]
\centering
\input{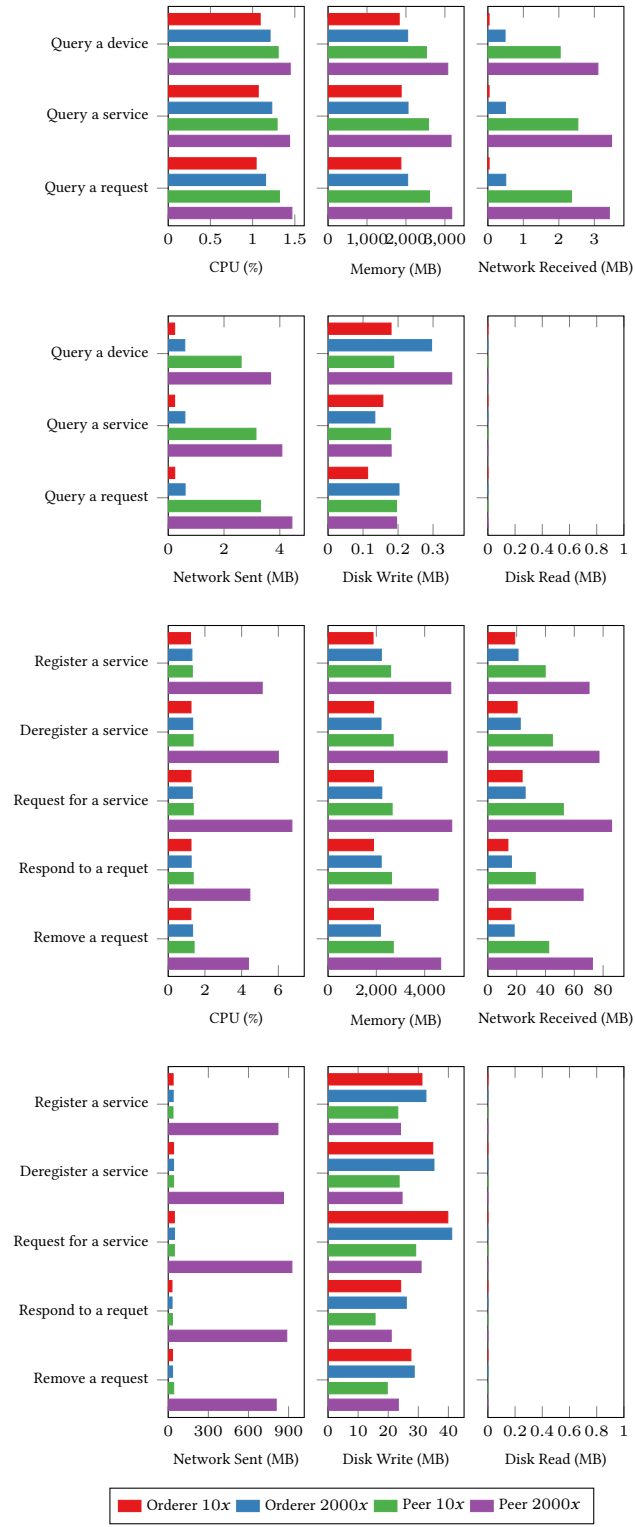}
\caption{Average system resource utilization of orderer and peer nodes for 10 client connections ($10x$) and 2,000 client connections ($2,000x$).\label{fig:utilization-vs-connection}}
\end{figure}

The system resource utilization also rises in the 2,000-connection scenario, as shown in Figure~\ref{fig:utilization-vs-connection}.
The most significant increases in relative resource utilization happen in networking, where we see a hundred-time growth in outgoing network traffic.
We also see a drastic usage increase in CPU and memory for both read and transaction operations.
Additionally, transaction operations are more fickle to changes in connection size compared to read operations, and so do peer nodes than orderer nodes.
Disk usage, however, does not vary too much with connection size.
The overhead observed in system resource utilization could be caused by connection overhead because the nodes have to maintain the connection from each client, verify client identities, and secure mutual communications using encryption.
The results also imply that the memory and network can quickly become the bottleneck of the blockchain platform for large IoT networks.

\subsection{Security and Privacy}

It is essential that our proposed platform provides a secure environment for IoT devices and applications leveraging blockchain technologies and modern cryptography.
It is also important that we preserve user and data privacy wherever needed.
In this section, we evaluate the security and privacy of our proposed system from the blockchain, auxiliary storage, and IoT device perspectives.

1) \textit{Blockchain security and privacy.}
\citeauthor{ferrag_blockchain_2018} \cite{ferrag_blockchain_2018} presented a comprehensive view of the thread models and attacks against blockchain systems.
They classify attacks on blockchain systems into five categories: identity-based attacks, manipulation-based attacks, cryptanalytic attacks, reputation-based attacks, and service-based attacks.
Recent blockchain systems are designed with these attacks in mind, and most of the attacks can be defeated if the system is configured following best practices.
It should also be noted that the design of permissioned consortium blockchains inherently makes many attacks more difficult. 
For instance, most blockchains use unique transaction IDs and nonces to protect transactions from replay attacks.
Permissioned consortium blockchains make Sybil attacks difficult to conduct as identity management is limited to organization administrators.

Regarding our proposed platform, its blockchain security lies within the design and implementation of Hyperledger Fabric.
\citeauthor{brotsis_security_2020} \cite{brotsis_security_2020} highlighted four attack surfaces of Hyperledger Fabric, namely consensus, chaincode, network, and privacy-preserving mechanisms.
Hyperledger Fabric is protected against most consensus-oriented attacks but is more vulnerable to non-deterministic behaviors in chaincode implementations and compromised participants.
While the latter threat can be eased with considerate deployment and maintenance, we carefully design and implement the chaincodes of our proposed platform to eliminate non-deterministic behavior and ensure the consistency of transactional data.
For example, the chaincodes use deterministic JavaScript Object Notation (JSON) serialization libraries to format results.
Also, the application generates all timestamps in the transactions instead of creating them when chaincodes are executing on peers.
This way, we can eliminate failed transactions due to the system clock not synchronizing across peer nodes.
Finally, chaincodes always check the caller's identity and input parameters to prevent impersonation attacks and invalid requests.

As for privacy, the identities and transactions are visible to all consortium participants.
Although this is usually expected in a trusted environment, the users of our proposed platform have the option to conceal the IoT data with the help of auxiliary storage and encryption.
Also, the private data collection feature offered by Hyperledger Fabric and zero-knowledge proofs \cite{kang_fabzk_2019} are promising technologies that can improve identity and data privacy in our platform.
We leave this as an avenue for future work.

2) \textit{Auxiliary storage security and privacy.}
We evaluate the security and privacy of the auxiliary storage systems using the confidentiality, integrity, and availability (CIA) model.
Confidentiality means that the IoT data in the storage should be accessible only to authorized users.
In our proposed system, securely passing sensitive data between a service provider and consumer can be achieved using a one-time encryption key or access token in terms of data streams.
A key may be asymmetrically encrypted using the receiver's identity and sent via the blockchain.
Regarding confidentiality during data transfer, the blockchain and auxiliary storage enforce encryption through Transport Layer Security (TLS).
Integrity ensures that unauthorized users do not alter data.
The data may be modified accidentally due to system errors or by a malicious party.
Our proposed platform utilizes blockchain as a layer of data integrity assurance for data in auxiliary storage because of its immutability.
Therefore, data providers are encouraged to include a digital digest and signature of the data alongside URI in the service requests and responses.
Finally, availability measures how often the data is accessible to its users.
For IoT data that desire a high level of availability, distributed storage such as IPFS may be used to facilitate data dissemination.
We leave the flexibility of ensuring data availability to the consortium administrators and application developers.

3) \textit{IoT device security and privacy.}
IoT devices have a long history of being a weak link to IoT system security.
Apart from being exposed to physical attacks such as node capturing, sleep deprivation, and false-data injection\cite{hassija_survey_2019}, IoT devices are also vulnerable to network-based attacks.
For example, insecurely configured devices are often targeted by IoT botnets.
An attacker can acquire access to such devices by brute-forcing login credentials or exploiting software flaws.
It then injects malware into these devices to grow the zombie network or initiate DDoS attacks against other targets.

The proposed IoT service platform can remedy network-based attacks against IoT devices.
It offers a sole secure communication channel to the IoT network that can replace insecure communication protocols such as Telnet and HTTP.
The attack surface of IoT devices shrinks as the number of needed services decreases.
The proposed platform also eliminates the need for weak credentials by employing strong cryptography keys and certificates.
It also enables automatic decommission of old IoT devices or decommission of compromised devices using certificate revocation mechanisms.
Finally, the use of blockchain and decentralized storage also enhances system security due to the absence of a centralized server, which is often the SPOF in IoT systems.

\section{Conclusion}
\label{sec:conclusion}

This paper has presented an innovative platform for secure and decentralized IoT communications utilizing the consortium blockchain.
The proposed platform models IoT communications as services supported by smart contracts.
The service provider, usually an IoT device, exchanges messages with service users securely through blockchain transactions.
To support a wide range of application scenarios, the platform also incorporates an auxiliary storage system as a secondary communication channel whose data integrity can be ensured by the blockchain.
Meanwhile, the inclusion of platform SDKs and gateway ease the complexity of integrating the proposed platform into existing IoT systems and devices.
Furthermore, we present a prototype implementation as well as exemplary applications to showcase the proposed platform's generality and versatility.
This paper later elaborates the experimental setup, methodology, and metrics we used to evaluate the performance of our solution.
Since the performance of a blockchain system is influenced by a variety of factors, we measure the platform's transaction throughput, latency, and the hardware resource utilization under different blockchain configurations and connection sizes.
The results indicate that the performance of read operations primarily depends on message size, while the transaction operation performance is subject to batch size, batch timeout, and connection size.
Nevertheless, our proof-of-concept implementation can achieve a throughput of 800 RPS and latency of 50ms for read transactions, and a throughput of 80 TPS and latency of 1s for write transactions when the blockchain parameters are optimized.
Overall, our proposed work shows great performance and usability potential as a blockchain-based secure communication platform for IoT.

The future work will focus on improving transaction throughput and latency for transaction operations on the platform using state-of-art lightweight consensus algorithms.
Additionally, we plan to investigate new approaches that integrate auxiliary storage with the blockchain to provide the same level of data security and integrity as blockchain transactions.
Lastly, we will explore new ideas to address privacy concerns and support private services.

\begin{acks}
The authors acknowledge the University of Tennessee at Chattanooga and Chameleon Cloud for providing experiment resources to this study.
\end{acks}

\bibliographystyle{ACM-Reference-Format}
\bibliography{main}

\end{document}